\documentclass[english, 11pt,a4paper]{article}
\usepackage{babel}
\usepackage[top=3cm, bottom=3cm, inner=3cm, outer=3cm]{geometry}
\usepackage[T1]{fontenc}
\usepackage[ansinew]{inputenc}
\usepackage{graphicx}
\usepackage{graphicx,amsmath,booktabs,xcolor,times,mathptmx}
\usepackage[font=small]{caption}	
\usepackage{natbib}
	\bibpunct{(}{)}{;}{a}{}{,}

\newenvironment{lcase}{\left \lbrace \begin{aligned}}{\end{aligned}\right.}

\begin{document}
\author{Giangiacomo Bravo\thanks{Dipartimento di Scienze Sociali, Università di Torino and Collegio Carlo Alberto.}{ } \& Lucia Tamburino\thanks{Dipartimento Territorio e Sistemi Agro-Forestali, Università di Padova.}}
\title{Are two resources really better than one? Some unexpected results of the availability of substitutes}%
\date{}
\maketitle

\begin{abstract}
The possibility of exploiting multiple resources is usually regarded as positive from both the economic and the environmental point of view.  However, resource switching may also lead to unsustainable growth and, ultimately, to an equilibrium condition 
which is worse than the one that could have been achieved with a single resource.  We applied a system dynamics model where users exploit multiple resources and have different levels of preference among them.  In this setting, exploiting multiple resources leads to worse outcomes than the single-resource case under a wide range of parameter configurations.  Our arguments are illustrated using two empirical situations, namely oil drilling in the North Sea and whale hunting in the Antarctic.

\medskip
\emph{Keywords: resource management; paradox of enrichment; environmental sustainability}
\end{abstract}

\section{Introduction}

Both commonsense and scientific inquiry suggests that exploiting two (or more) resources is both economically more profitable and reduces depletion risks in comparison with the case where only one resource is available.  To a large extent, the possibility of exploiting new resources represents also one of the strongholds for the idea of economic growth itself.  For instance, the well known critique of \citet{Beckerman1992} of the idea that resource constraint represents a limit for economic  growth was based on the effects that economic feedback mechanisms have on the development of new resources, i.e, on `` [\ldots] the various effects of a rise in the price of any resource that is becoming scarce: increased profitability of exploration, of improved processing techniques, of \emph{increased research into substitutes}; \emph{increased replacement of the resource by existing substitutes in final use}; [\ldots]'' \citep[483, emphasis added]{Beckerman1992}.

The fact that relying on multiple resource is beneficial for the stability of the system is a common statement also in community ecology.  For instance, switching preys is known to help to sustain ecosystems by allowing a recovering of depleted populations \citep[e.g.][]{Ajraldi2008,Khan2004,Matsuda1986}.  This result is not limited to natural systems.  For instance, \citet{Katsukawa2003} showed  that switching among different targets improves both yields and sustainability for fisheries under a wide range of conditions.  

However, in the real world, larger resources often imply a faster growth of the exploiting population or industry.  This may produce a positive feedback that actually \emph{accelerates} resource depletion in comparison with a case where less abundant resources are available.  This is not limited to human activities.  The issue has indeed a long tradition in ecology, where it is often called the \emph{paradox of enrichment} \citep{Rosenzweig1971}.  According to the paradox argument, the increase of the carrying capacity of a resource population leads to a destabilization of a previously stable system, involving a risk of extinction for both the resource and the exploiter populations.  The idea of an enrichment paradox was subsequently contradicted by experimental results \citep{Kirk1998,McCauley1990,Persson2001}, while \citet{Kuang1999} proved mathematically that the paradox cannot occur under the empirically relevant condition of ratio-dependent systems: i.e., systems 
where  the consumption rate depends on the ratio between the resource and the exploiter populations.\footnote{More formally, $x$ being the resource population and $y$ the exploiter one, a system is ratio-dependent if resource consumption is given by $y f(x/y)$, where $f$ represents a generic function.}

Nevertheless, recent research provides new support for Rosenzweig's idea.  \citet{Chatterjee2008} showed that the paradox can occur also in ratio-dependent systems, when further factors are included in the model to make it more realistic, e.g., when the resource population can be affected by a transmissible disease.  Moreover, \citet{Tamburino2010} developed a mathematical model to study the interactions of wild and domestic herbivores with a natural or man-maintained landscape.  Their argument is that, when herbivores rely on two different resources---i.e., grass and young trees---the growth of their population produces higher environmental damage than when only grass is available or consumed.  In the long term, this can also lead to a sudden decline of the herbivore population itself.   It is worth noting that \citet{Tamburino2010} also applied their model to two different case studies, hence providing empirical support to the theoretical argument.

In this paper, we explore a different case of the enrichment paradox by applying this idea to situations where multiple resources are simultaneously available.  This can be regarded as an extension of Rosenzweig's model
where the enrichment is given not by an increase of the resource carrying capacity but by the availability of adequate substitutes.  More specifically, we further develop the \citet{Tamburino2010} model by generalizing their case and applying it to human management of natural resources.  The main difference with previous models is in the introduction of different preference levels  among resources.  In other words, our model takes into account that real-world resources are never exact substitutes for each other and that users usually have different levels of preferences between them.  Note also that the simultaneous presence of several substitutes permits of modeling the 
enrichment also when resources are non-renewable, hence making our model more general and more suited to applications to important real-world cases (see Section \ref{sec:oil}).   

Overall, our findings suggest that relying on multiple resources can lead to an unsustainable growth of the exploiting industry that produces worse outcomes than the single-resource exploitation case under a wide range of conditions.  The argument is supported by two empirical applications.  First a non-renewable resource exploitation case is studied: oil production in the Norwegian North Sea.   Our analysis showed that the exploitation of small fields, besides the giant reservoirs originally discovered, actually accelerated the overall depletion rate of oil reserves.  Our second case regards pelagic whaling in the Antarctic, where more than fifty years of hunting relied on target shifting from the most to, progressively, the least productive (i.e., smaller) whale species.  We showed that, if hunters had targeted only the largest species, this not only would have avoided the commercial extinction of other whales, but it would have reduced the impact on the main targets as well.  Moreover, this would have avoided the dramatic drop of the whaling industry that occurred during the 1960s and 1970s, permitting a more sustainable management of the resource also from the economic point of view.

This paper is organized as follows.  Section \ref{sec:model} presents our model and derives some general indications in an abstract context.  Section \ref{sec:oil} applies the model to the Norwegian oil case.  Section \ref{sec:whales} analyzes whale hunting.  Finally, Section \ref{sec:conclusions} discusses the general implications of our results.

\section{The general model}\label{sec:model}

The system developed by \citet{Tamburino2010}  included a herbivore population and its resources and it can be regarded as a predator--prey system with two resource populations, namely grass and trees.   Even if herbivores consume mainly grass, they occasionally debark trees.  This occurs especially in the winter season, when grass availability is low \citep{Sharrow1992}.  The authors applied their model to herbivores grazing in both urban and natural parks.  The same model, with relatively small modifications, is also able to represent a generic system with a human population exploiting a variable number of resources.  Generalizing its form, the two resource model developed by Tamburino and Venturino becomes:
\begin{equation} 
\begin{lcase}
	\dot{U}=&-\mu U + e_1 \frac{UR_1}{c_1+U+\alpha _1 R_1} + e_2 \frac{UR_2}{c_2+U+\alpha _1  R_1+\alpha _2 R_2} \\
	\dot{R_1}=& f_1(R_1)- \frac{UR_1}{c_1+U+\alpha _1 R_1}\\
	\dot{R_2}=& f_2(R_2)- \frac{UR_2}{c_2+U+\alpha _1 R_1+\alpha _2 R_2}
\end{lcase}
\label{mod-original}
\end{equation}
where $U$ refers to the users' population, while $R_1$ and $R_2$ refer to the exploited resources.  While in the original model $\mu$ represents the metabolic rate of herbivores and $e_1, e_2$ their assimilation coefficients, we consider them respectively as the users' capital depreciation and the proportion of the income deriving from the exploitation of each resource that is reinvested in the same activity.  

In \citet{Tamburino2010}, the resource growth function, here generically indicated with $f_j$,  is the logistic $f_j=r_jR_j(1-R_j/K_j)$, where $r_j$ represents the growth rate of $R_j$ and $K_j$ its carrying capacity, for $j=1,2$.  In our generalization, $f_i$ can instead be any  growth function of the resources under examination, e.g.  it can model inter-specific competition among the different resource populations or can be set to zero in the case of non-renewable resources.

The function modeling users' consumption of the first resource is given by the term
\begin{displaymath}
\frac{UR_1}{c_1 + U+ \alpha _1R_1}
\end{displaymath}
where $\alpha _1$ and $c_1$ denote positive constants.  This function  follows the Michaelis-Menten model \citep{Murray1989}, namely:
\begin{displaymath}
	\lim_{U\rightarrow \infty }\frac{UR_1}{c_1 + U+ \alpha _1R_1}=R_1 \hspace{1 cm} \textrm{and} \hspace{1 cm}
	\lim_{R_1\rightarrow \infty }\frac{UR_1}{c_1 + U+\alpha _1R_1}=\frac{U}{\alpha _1 }
\end{displaymath}
meaning that when $U$ grows, the consumption can be at most the whole amount of the first resource in the system and, when the availability of first resource is unlimited,  its consumption  has an upper bound proportional to the number of users: $\alpha _1^{-1}U$.  These are reasonable results, making this function biologically suitable.  From the second property, we deduce that $\alpha_1^{-1}$ represents the maximum consumption per user's unit in the presence of an unlimited availability of the first resource.  The parameter $c_1$, called the Michaelis constant, or sometimes the half saturation constant, affects instead only the speed at which consumption approaches its asymptotes.  

The function modeling the consumption of the second resource is similar but, following the ideas of feeding switching proposed in the classical \citet{Tanski1978} paper and in a number of more recent ones \citep[e.g.][]{Ajraldi2008,Bean2006,Khan2002,Khan2004, Vilcarromero2001}, the first resource appears in the denominator as well.  This function models a different level of users' preference for each of the resources.  For instance, herbivores in \citet{Tamburino2010} consumed mainly grass and only occasionally switched their attention to trees, with an increasing rate when the first resource becomes scarce.  Actually this is a rather general situation:  when several resources are available, they often present slightly different characters, making each of them only an imperfect replacement of the others.  Usually some are indeed preferred, due to easier access and/or higher marginal value.  For instance, traditional agricultural populations preferentially cultivated plains, moving to the hillsides only after a depletion of the fields or due to population pressures \citep[e.g.][]{Diamond2005,Ponting1991}.  Similarly, surface water is usually exploited before groundwater and renewable groundwater is preferred to fossil aquifers or de-salinization \citep[e.g.][]{Al-Rashed2000}.

Model (\ref{mod-original}) is hence well suited to represent a generic system where a population of users  exploits two or more resources with different preference levels.  Extending (\ref{mod-gen}) to the $n$ resource case $\{R_1, \dots , R_n\}$, with resources ordered by decreasing preference levels, we obtain
\begin{equation} 
\begin{lcase}
	\dot{U}=&-\mu U + e_1 \frac{UR_1}{c_1+U+\alpha _1 R_1} + e_2 \frac{UR_2}{c_2+U+\alpha _1 R_1 + \alpha _2R_2}+\\
	&  + \dots + e_n \frac{UR_n}{c_n+U+\Sigma _{i=1}^n \alpha _i R_i} \\
	&  \dots \\
	\dot{R_j}=& f_j(R_j)- \frac{UR_j}{c_j+U+\Sigma _{i=1}^j \alpha _i R_i}\\
	&  \dots \\
\end{lcase}
\label{mod-gen}
\end{equation}

Let us recall that $\alpha_i^{-1}$ represents the consumption per unit of user in the presence of an unlimited availability of the $i$-\textit{th} resource, namely the amount of resource a unit of user extracts from the system.  Actually, the user's income is only the proportion of this amount  given by $e_i / \alpha_i$.  From this consideration, it is easy to see that the following relation must hold
\begin{equation}
\frac{e_i}{\alpha _i} > \frac{e_j}{\alpha _j} \hspace{1 cm} \forall i > j
\label{eq:e-alpha}
\end{equation}
which simply means that the income which users are able to obtain from a given resource is greater than the one obtainable from any of the following resources, making the first resource indeed preferred.  Moreover,  from (\ref{mod-gen}) we deduce that the $i$\textsuperscript{th} resource is able to sustain the users' population by itself if and only if the assimilated part of the amount consumed, namely $e_i/\alpha_i$, is able to compensate for capital depreciation, i.e., 
\begin{equation}
 \frac{e_i}{\alpha _i}>\mu 
 \label{enough}
\end{equation}
>From (\ref{enough}), we also deduce that there is a strict relation between the parameters $e_i$, $\alpha_i$ and $\mu$, with an increase of, for instance, $\mu$ producing similar outcomes as an increase of $\alpha_i$ or a decrease of $e_i$.  

To study the model's behavior, we ran numerical simulations of the system  (\ref{mod-original}) using three resources.  As in \citet{Tamburino2010}, we employed a logistic growth function for all the resources.  Being interested in comparing the case where users exploited only one resource with multiple-resource exploitation, we ran  simulations  using the same parameter values, but varying the number of exploited resources, namely three, two, or only one resource.  These different exploitation strategies were examined under two different scenarios.  In the first one, each resource was able  to sustain users by itself, which means that relation (\ref{enough}) holds for every $i$.  In the second scenario, the same relation holds only for $i=1$.
Note that if this assumption did not hold, users could not survive without secondary resources, making a comparison between single and multiple resource exploitation of little interest.

\begin{figure}[!t]
\centering
\includegraphics[width=0.333\textwidth]{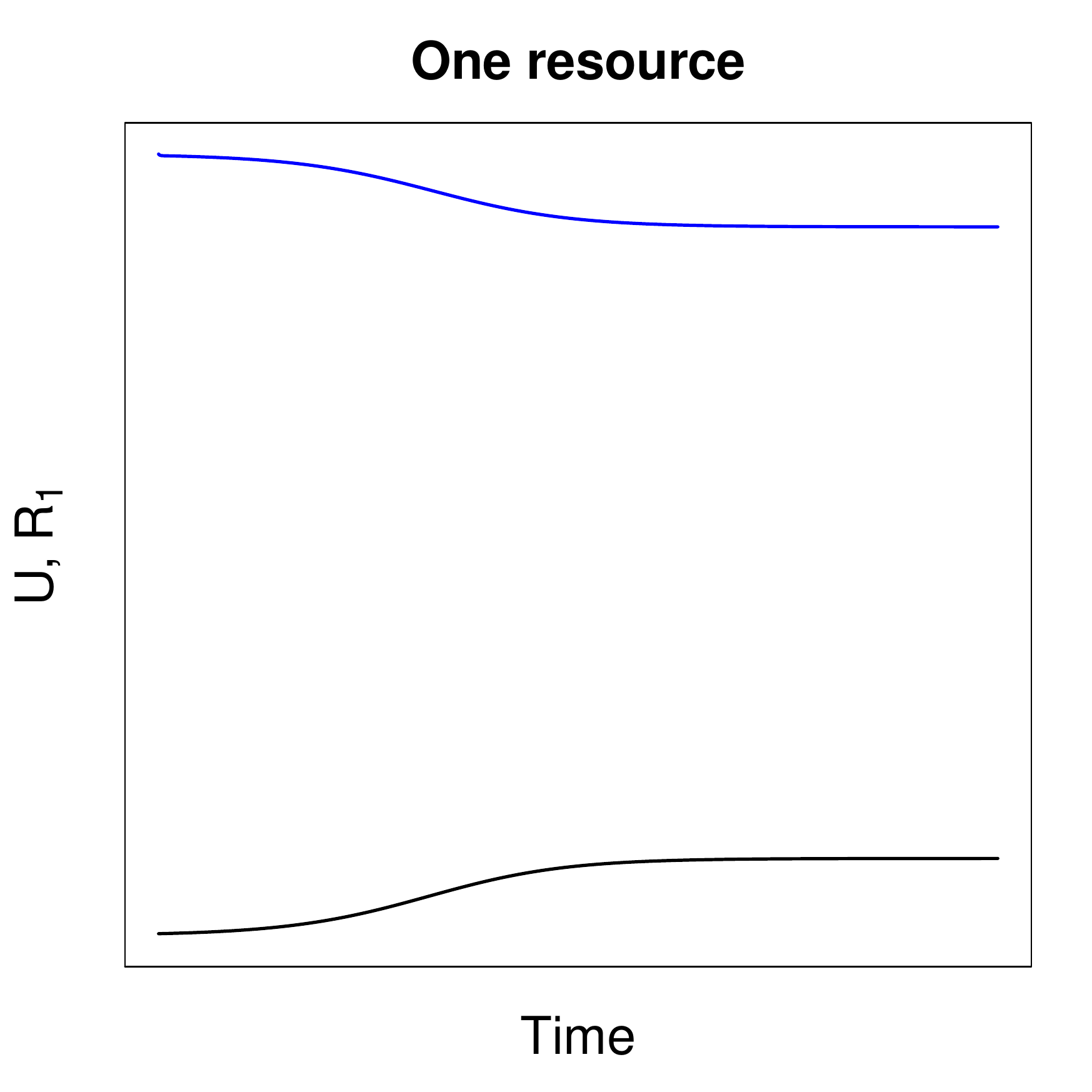}\includegraphics[width=0.333\textwidth]{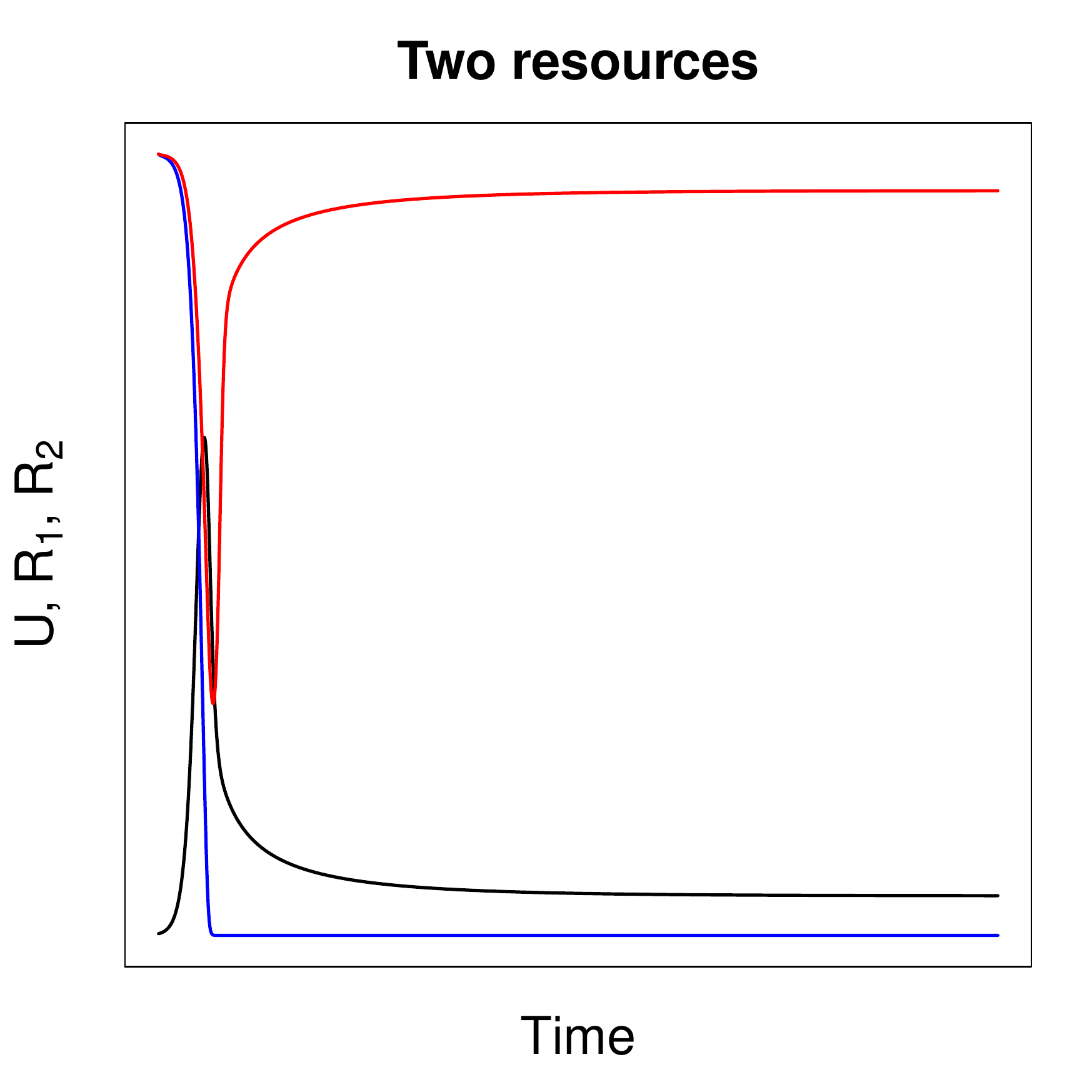}\includegraphics[width=0.333\textwidth]{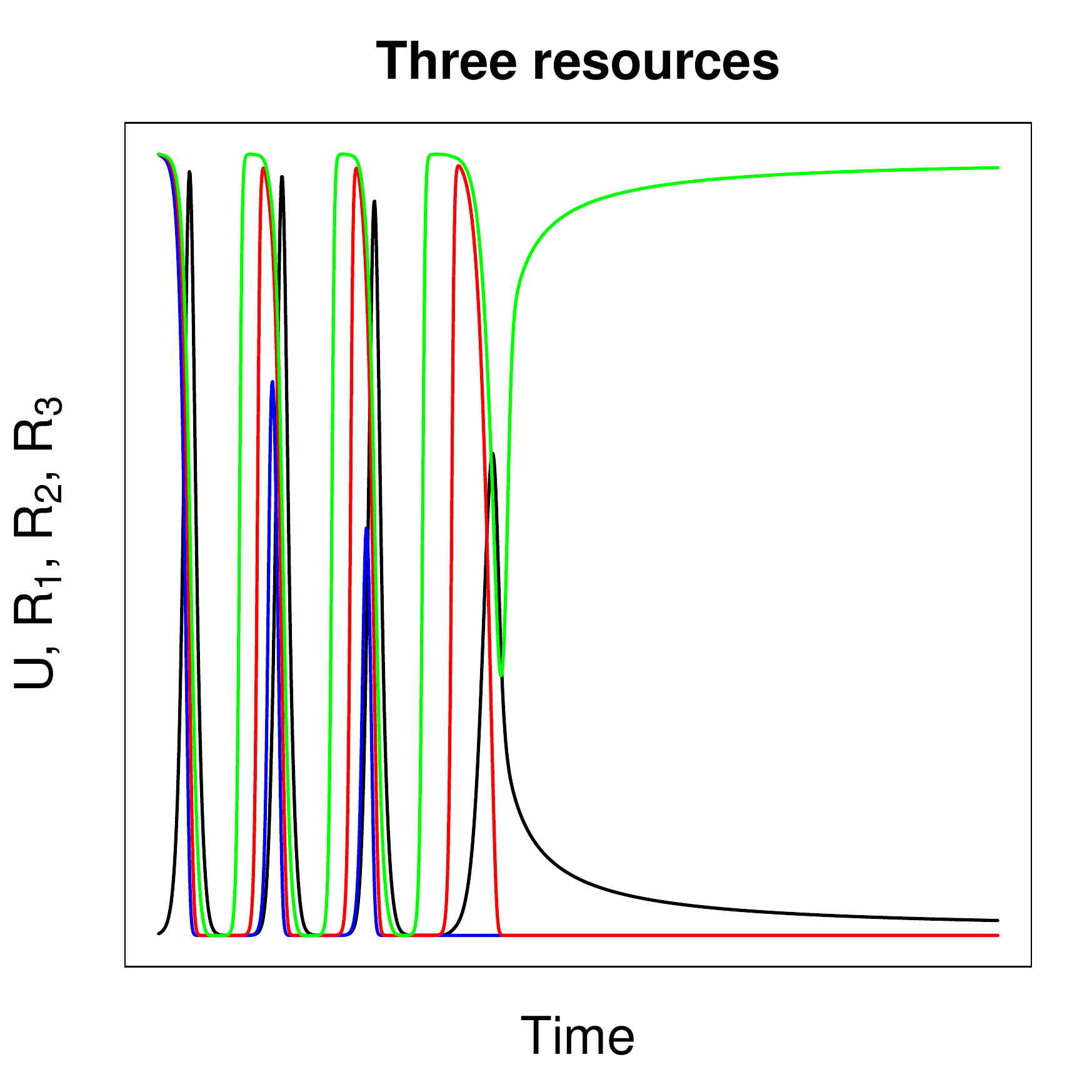}\\
\includegraphics[width=0.333\textwidth]{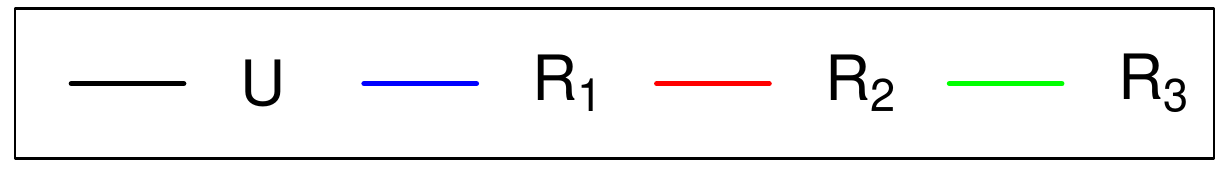}
\caption{Model dynamics for the three possible exploitation strategies under the first scenario, where all resources are able to sustain users by themselves.  The resource growth function follows a logistic scheme $f_j=r_jR_j(1-R_j/K_j)$.  Parameters for the plotted models: $\mu=0.2$, $\alpha_1=\alpha_2=\alpha_3=1$, $K_1=K_2=K_3=1000$, $c_1=c_2=c_3=1$, $r_1=r_2=r_3=0.5$, $e_1=0.21$, $e_2=0.205$, $e_3=0.201$.  Initial conditions: all resources are at carrying capacity; one user enters the system.}%
\label{generic1}%
\end{figure}

Figure  \ref{generic1}  presents the results for the different strategies under the first scenario.  With the chosen parameters, users  reach an equilibrium independently of their exploitation strategy but, in the case of a multiple-resource exploitation, an equilibrium is attained only with the least preferred resource, after destroying  the favorite one(s).  As a consequence, users' equilibrium levels slightly decreased when moving from one to three-resource exploitation.  Moreover, in the multiple-resource exploitation case, the system showed large oscillations while these did not occur in the one-resource case.   In this scenario, exploiting only one resource hence implied better results for both users and the environment: for users because a higher equilibrium was established and large oscillations were avoided, for the environment because no resource reached full depletion.  

\begin{figure}[!t]
\centering
\includegraphics[width=0.333\textwidth]{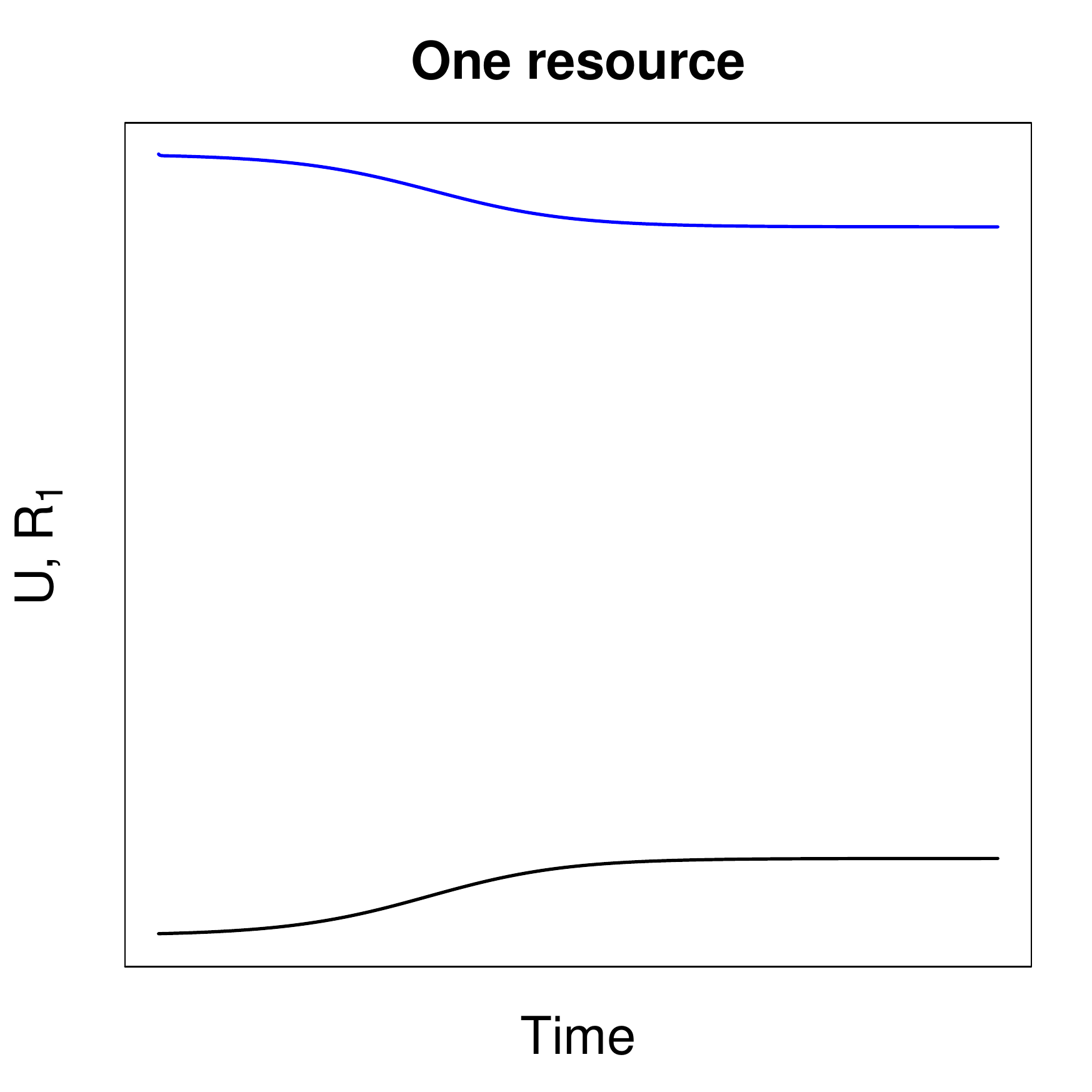}\includegraphics[width=0.333\textwidth]{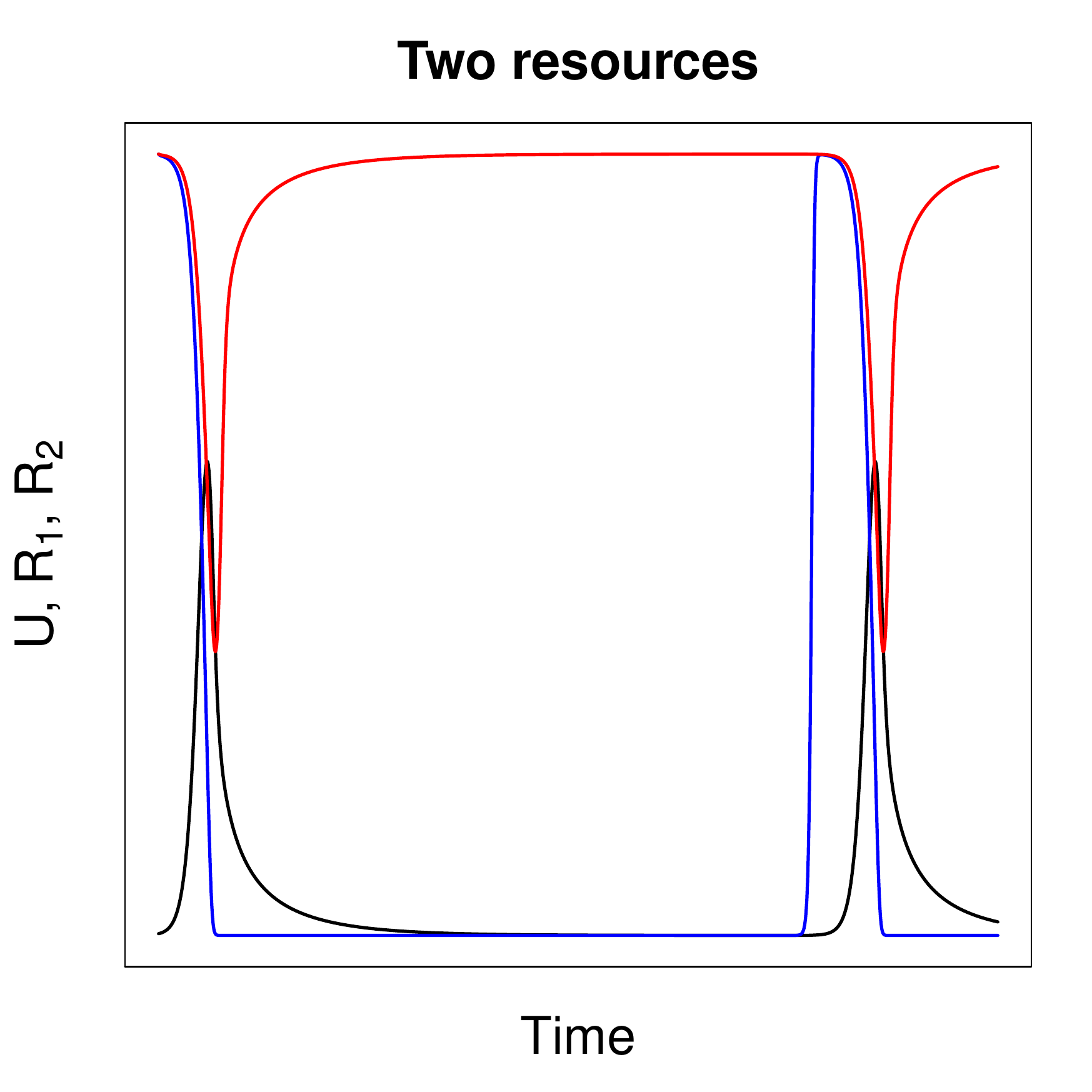}\includegraphics[width=0.333\textwidth]{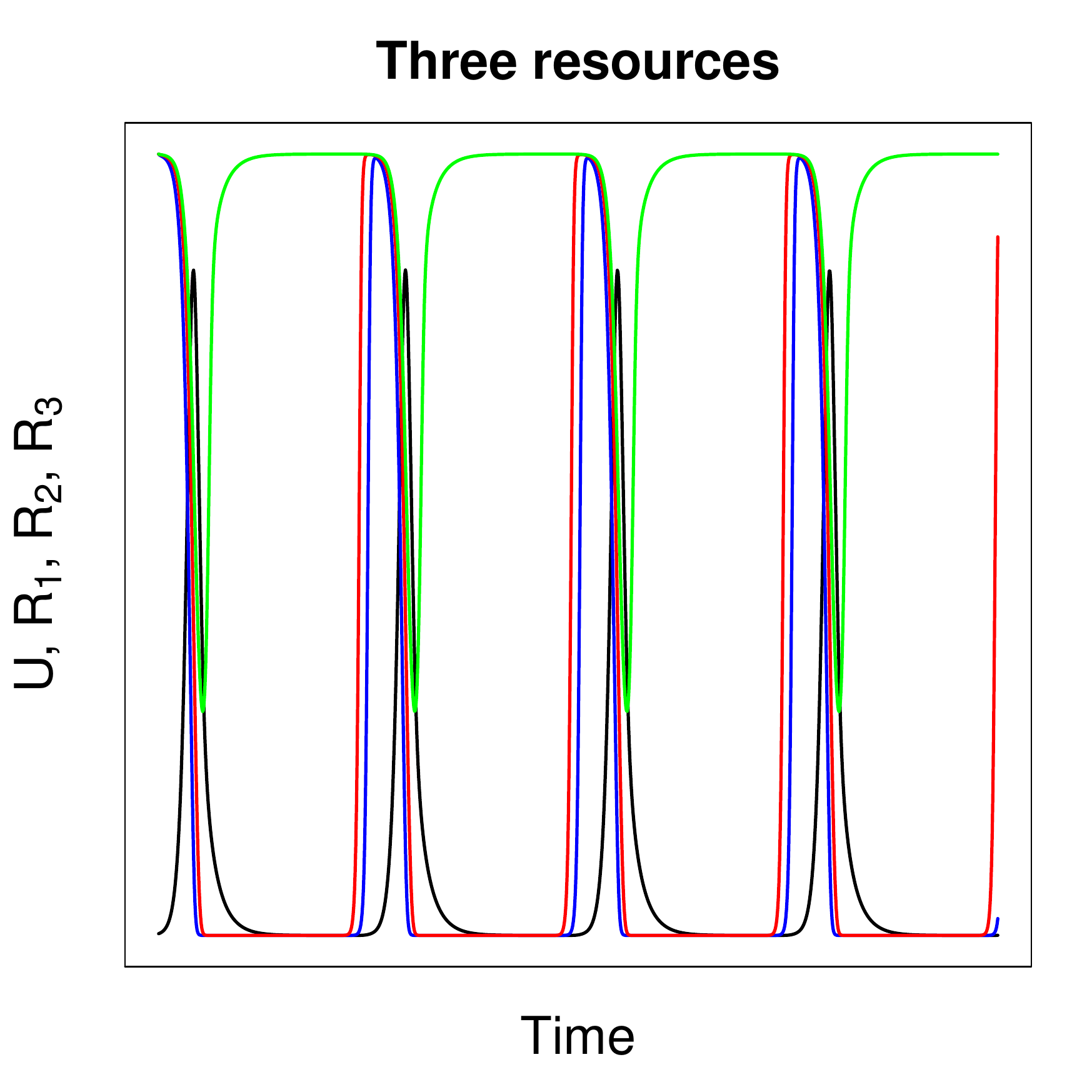}\\
\includegraphics[width=0.333\textwidth]{legend}
\caption{Model dynamics for the three possible exploitation strategies under the second scenario, where only the first resource can sustain users alone.  The resource growth function follows a logistic scheme $f_j=r_jR_j(1-R_j/K_j)$.  Parameters for the plotted models: $\mu=0.2$, $\alpha_1=\alpha_2=\alpha_3=1$, $K_1=K_2=K_3=1000$, $c_1=c_2=c_3=1$, $r_1=r_2=r_3=0.5$, $e_1=0.21$, $e_2=0.19$, $e_3=0.15$.  Initial conditions: all resources are at carrying capacity; one user enters the system.}
\label{generic2}
\end{figure}

In the second scenario, when users exploit only one resource, they survive and find an equilibrium with $R_1$.   No equilibrium is instead reached when users exploit  two or three resources   (Fig. \ref{generic2}).  In these cases, the system oscillates with sudden burst of the primary and secondary resources, followed by a rapid increase of users and a subsequent drop of all populations.  Between two bursts, only the least preferred resource goes to carrying capacity, while users and the preferred resource(s) remain very close to zero.  In other words, between two bursts all the resources except the least preferred one are depleted and the exploiting industry is completely dismantled.  Moreover, unlike the model, in the real world when a population reaches extremely low levels it becomes unable to recover, leading to a high risk of actual destruction for both users and resources.  To sum up, in both scenarios, exploitation of more than one resource results in a worse situation from both the economic and the ecological points of view.

We also explored the system response under a wide range of parameter variation, observing  that the outcomes presented above are quite general and do not depend on the choice of a particular set of parameter values.  Since, as highlighted above, the system is markedly sensitive to the ratio between the parameters $\alpha _i$, $\mu$, $e_i$, we produced Hopf bifurcation diagrams varying each of these while keeping the other parameters unchanged.  As an example, let us describe the diagram obtained by varying $e_1$ and setting the other parameters as in the second scenario.   For each value of $e_1$, we ran the system for 2000 time steps recording only the last 500 in a dataset.  We then plotted on the $y$-axis the minimum and the maximum values of $U$ (but the same could be done with $R_1$, $R_2$ and $R_3$) corresponding to each value of $e_1$.  When minima and maxima coincide, the system rests in a stable equilibrium, while a difference between the  minima and maxima implies an oscillation between these extremes.  

\begin{figure}[!t]
\centering
\includegraphics[width=0.333\textwidth]{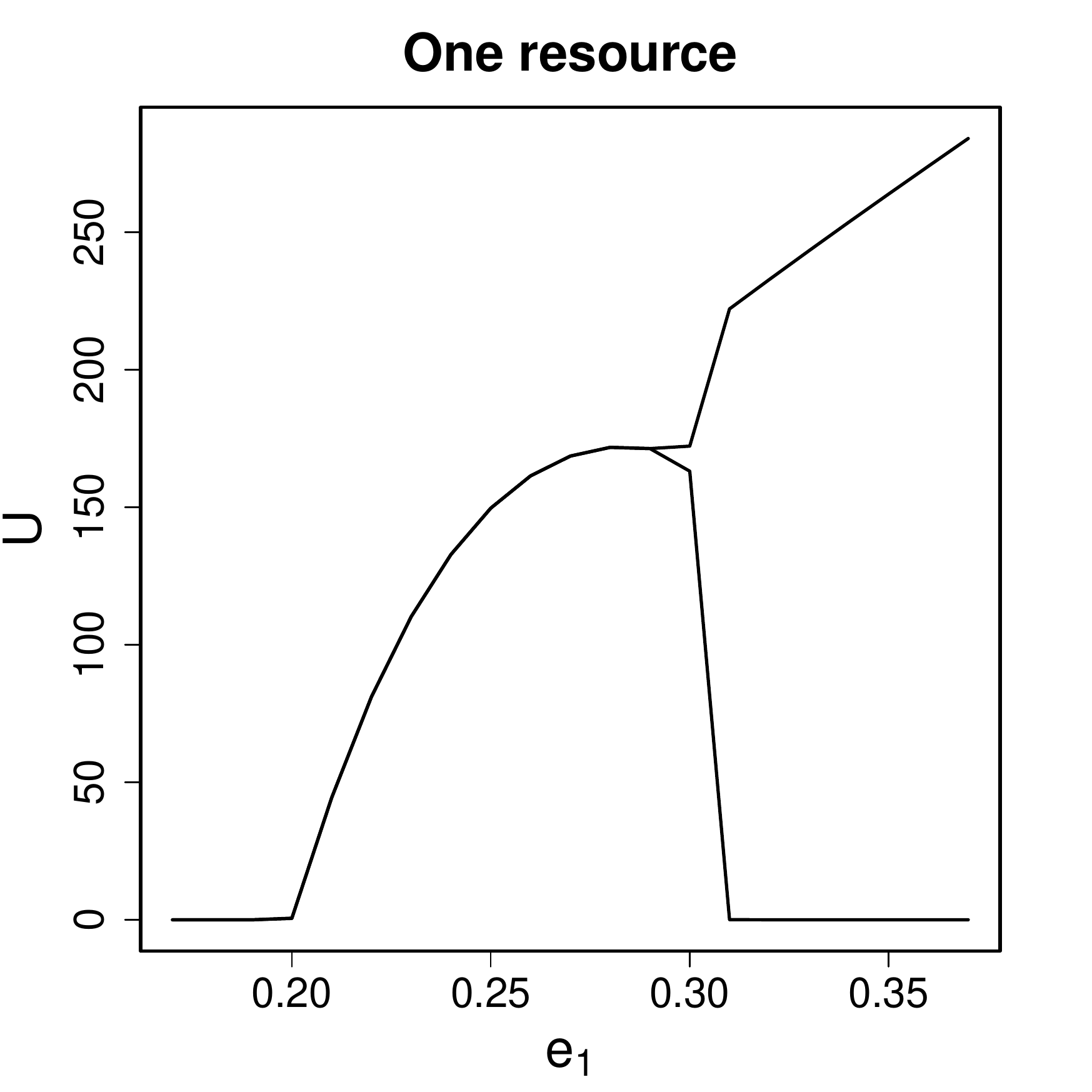}\includegraphics[width=0.333\textwidth]{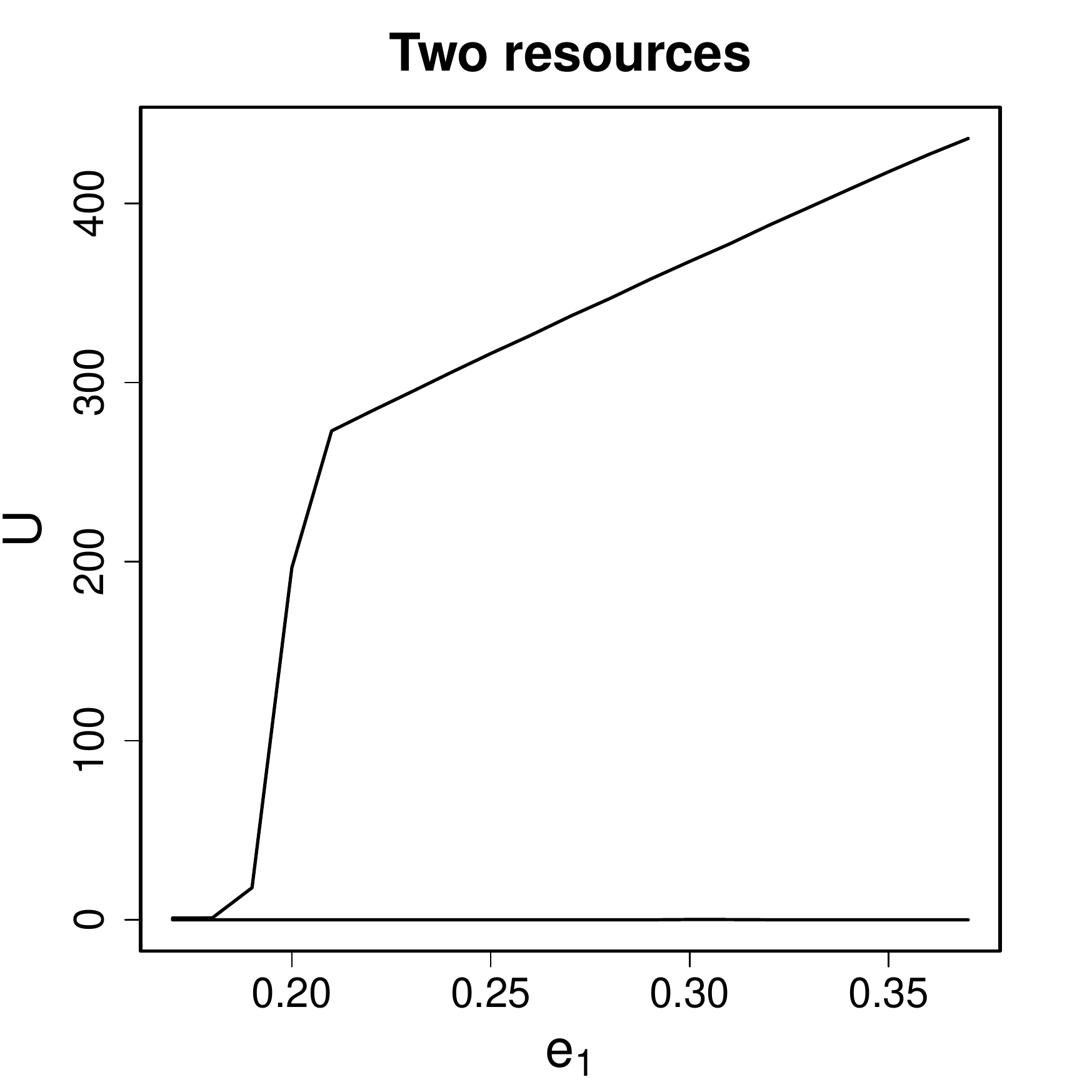}\includegraphics[width=0.333\textwidth]{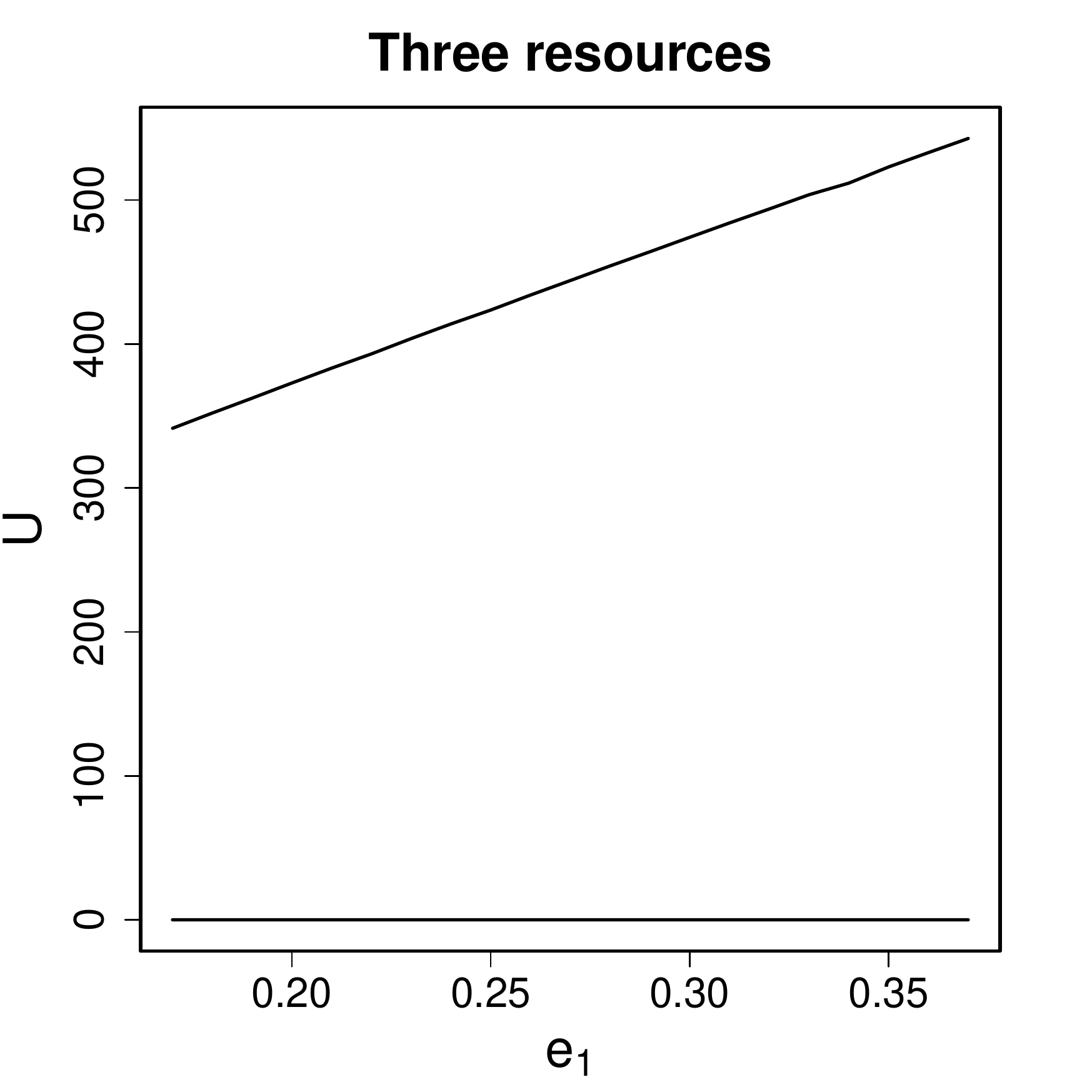}\\
\caption{Hopf bifurcation diagrams of users' population, varying $e_1$.  Other parameters were set as in the second scenario.  When exploiting only the favorite resource, an equilibrium is possible with values of $e_1$ up to about 0.30, while in the case of two or three resources there is no value range leading to an equilibrium which respect the assumption of a preferred resource able to sustain users alone.}
\label{hopf}
\end{figure}

The diagrams presented in Figure  \ref{hopf} show that, for values of $e_1$ above a certain threshold (depending on the other parameter values), the system starts to oscillate while below, it reaches a stable equilibrium.  This threshold varies depending on the number of resources exploited by users: the higher the number of resources the  more the threshold moved to the left.  This means that there is a whole range of values for $e_1$ where the equilibrium is possible only if users follow a single resource exploitation strategy, while exploiting multiple resources leads to an unstable system.   Oscillations become larger when the value of $e_1$ increases, with minima of both users and  resources (except the last one) almost reaching zero, exactly as in the case described above
 (Fig.  \ref{generic2}).  Similar results were obtained under a wide range of parameter configuration.

According to Figure  \ref{hopf}, multiple resources allow users to survive even with small values of $e_1$, while using a single resource leads $U$ to zero and multiple-resource exploitation would seem better.  However, in this case $e_1$ is below $0.2$ (or at least very close to it), which is the minimum value satisfying condition (\ref{enough}) for the first resource, given the current parameter setting.  This means that our assumption of a first resource's being able to sustain users alone no longer holds.  Vice-versa, as long as the assumption holds,  multiple-resource exploitation became less advantageous, at least in the explored  parameter ranges.   

To sum up, a multiple-resource exploitation strategy leads to the depletion of all resources except the least preferred one, under a wide range of parameter configurations.  Alternatively, this produces large oscillations, with all populations (except the last resource) almost vanishing at their minima.  Under the same parameter configurations, exploiting only $R_1$ does not produce such a complete depletion and instead leads to a stable equilibrium.  This implies the counter-intuitive idea that dependence on just one resource is better than having multiple exploitation options.

\section{Model application 1: Norwegian oil production}\label{sec:oil}

\begin{figure}[!t]%
\centering
\includegraphics[width=0.333\textwidth]{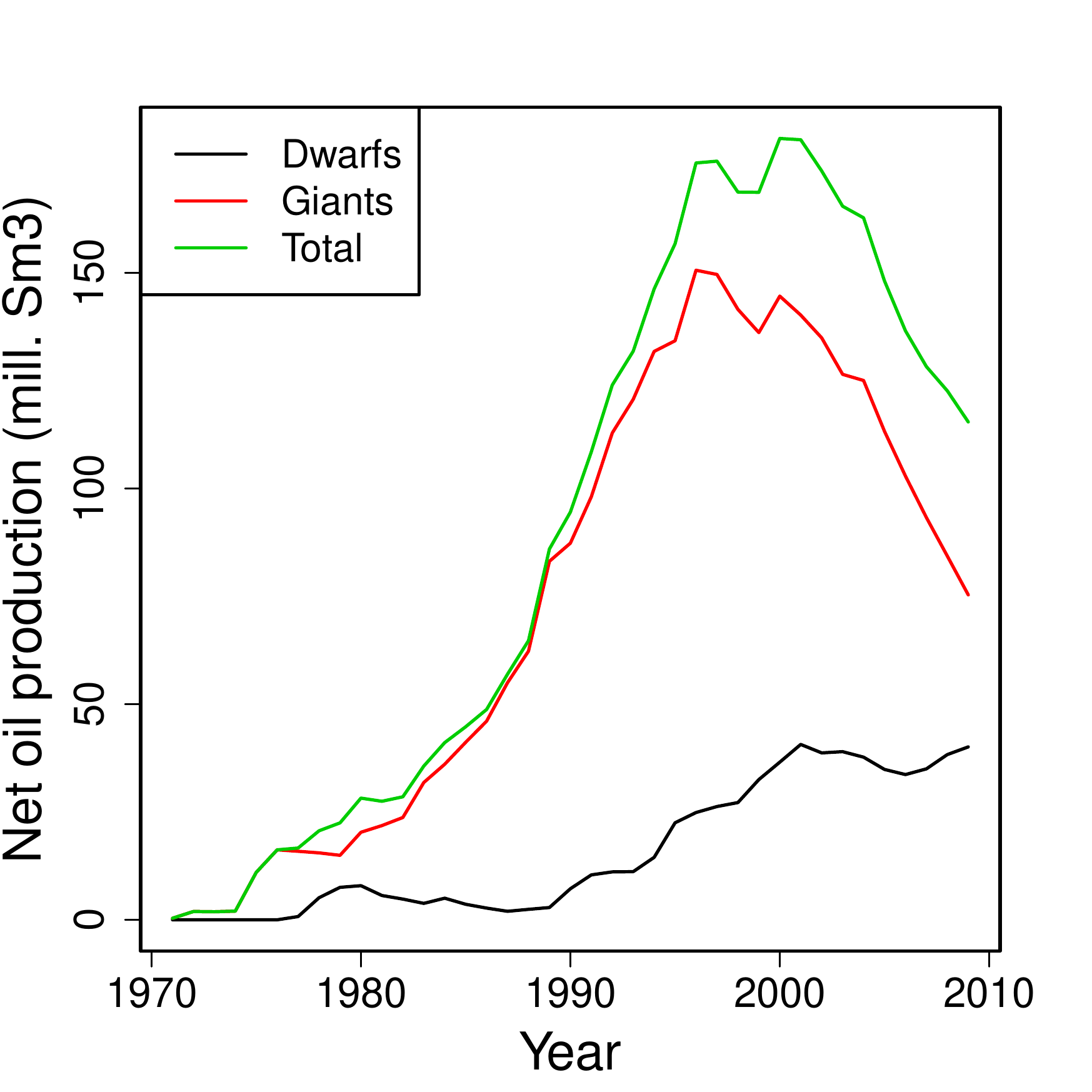}%
\caption{Oil production on the Norwegian continental shelf 1971--2009.  Source: Norwegian Petroleum Directorate's Fact-Pages (www.npd.no).}%
\label{fig:oilprod}%
\end{figure}

As a first application for our model, we choose the dynamics of oil production in the Norwegian continental shelf.  The case is especially interesting since Norwegian oil production reached a peak in 2001 and subsequently declined \citep{NPD2009}, representing hence a clear example of Hubbert's curve \citep{Hubbert1956}.\footnote{For a recent discussion of Hubbert's theory, see \citet{Brandt2007}.} Most of the Norwegian oil production came from a relatively small number of ``giant'' fields, while the larger number of ``dwarfs'' contributed only a minor part of the total output.  More specifically, \citet{Hook2008} identified 17 giant fields that, together, contributed more than 80\% of the total production.  Giant fields are easier to exploit and started their activity mostly in the early 1970s, while only since the middle of the 1990s did dwarfs begin to account for a significant share of production (Fig.\  \ref{fig:oilprod}).

In our model, we consider two resources, oil from giant fields and oil from dwarf fields, with the former being clearly the preferred one.  Both being  non-renewable, we have $f_i = 0 ~ \forall ~ i$.  There is also no need to distinguish between $e_1$ and $e_2$, since both parameters refer to the same kind of output (oil).  
The general model presented in (\ref{mod-gen}) becomes

\begin{equation} \label{eq:oil}
\begin{lcase}
	\dot{U}=&-\mu U + e \left(\frac{UR_1}{c_1+U+\alpha _1R_1} + \frac{UR_2}{c_2+U+\alpha _1R_1+\alpha _2R_2}\right) \\
	\dot{R_1}=&- \frac{UR_1}{c_1+U+\alpha _1R_1}\\
	\dot{R_2}=&- \frac{UR_2}{c_2+U+\alpha _1R_1+\alpha _2R_2}
\end{lcase}
\end{equation}

In (\ref{eq:oil}) the resources $R_1$ and $R_2$ represent oil reserves in, respectively, giant and dwarf fields, while users $U$ represent extraction companies.  Resources are expressed in terms of their value, with units corresponding to the value of one million standard cubic meters (Sm$^3$) of oil.  Users are expressed in the same units as capital embodied in extraction companies.  It is worth noting that, since $\mu U$ represents costs while the second and third terms of the users' equation represents revenues, $\dot{U}$ can be regarded as the net profit obtained by the oil industry in a given time unit.

Data regarding Norwegian oil reserves were extracted from the Norwegian Petroleum Directorate database and led to estimate $R_1(0) = 4016.4$ millions Sm$^3$ and $R_2(0) = 1241.6$ millions Sm$^3$ respectively.  Since we model the exploitation of Norwegian oil from its beginnings, in the early 1970s, we set $U(0) = 1$.

Recalling that the inverse of $\alpha_i$ represents the consumption of resource $i$ per unit of user (in condition of unlimited resource availability), $\alpha_i^{-1}$ can be viewed as the energy returns from oil extraction per unit of energy invested, a commonly used indicator called EROI (Energy Returns On Investments).  Recent research showed that EROI estimates for offshore oil drilling, including all direct and indirect energy costs, range in the $[5,10]$ interval \citep{Gately2007}.  We apply the upper limit of this estimate to  giant fields and the lower limit to dwarfs.  This leads to a value of $\alpha_1 = 0.10$ and of $\alpha_2 = 0.20$ respectively.  Estimations for $\mu$ and $e$ are based on financial data released from the Statoil company \citep{Statoil2009} and give $\mu=0.36$ and $e=0.74$.

The Michaelis constants $c_i$ affect the speed at which production approaches its asymptotes.  This is hard to derive from existing empirical data, since this would basically imply a forecast of future production trends, something that can be done only after the model calibration.  Nevertheless, these parameters can be estimated by fitting the model to the empirically observed dynamics.   Note that we employ data based on production figures, which are represented in our model by the second and third terms of the users' equation: $(UR_1)/(c_1+U+\alpha _1R_1)$ and $(UR_2)/(c_2+U+\alpha _1R_1+\alpha _2R_2)$ respectively.  
We hence choose the $c_i$ parameter combination minimizing the root mean square error (RMSE) between real and modeled production data.  This leads to an estimation of $c_1=5290$ and $c_2=9000$.

\begin{figure}[!t]%
\centering
\includegraphics[width=0.333\textwidth]{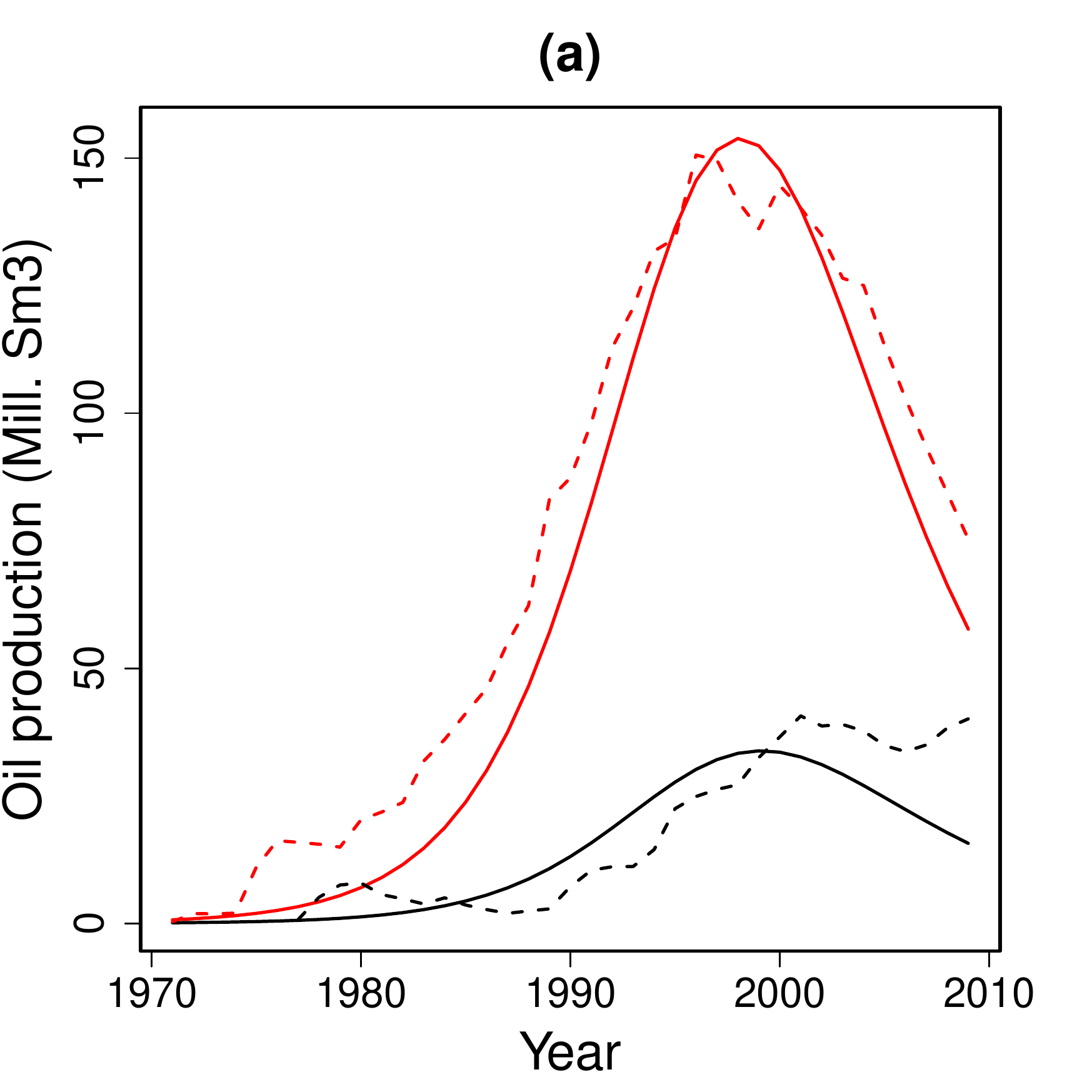}\hspace{1cm}\includegraphics[width=0.333\textwidth]{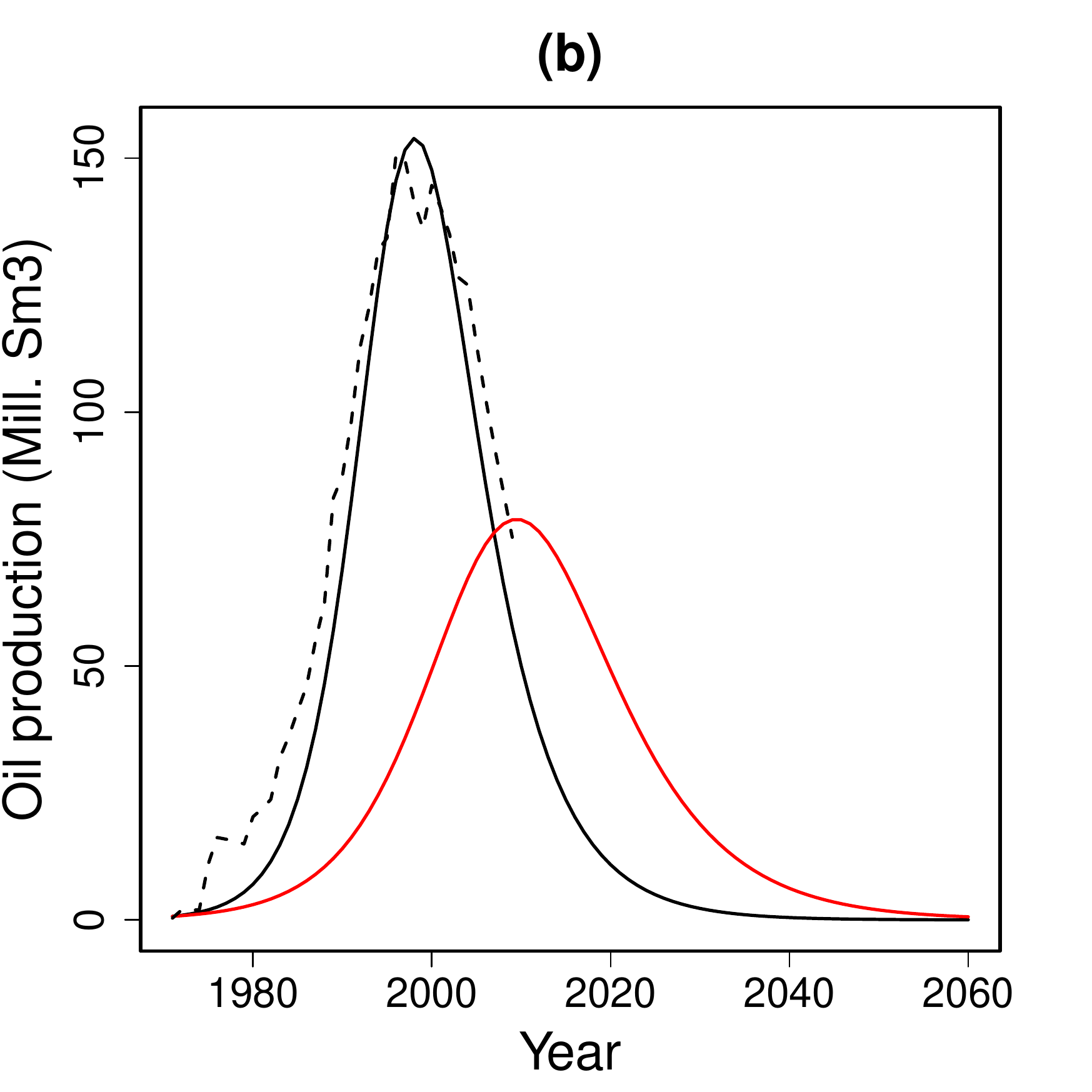}%
\caption{Model outputs.  (a) Oil production 1971--2009: model output (plain lines) and real data (dashed lines).  (b) Long term forecast of oil production from giant fields: including production from dwarfs (black line); excluding production from dwarfs (red line).}%
\label{fig:oilModel}%
\end{figure}

Using these parameters, our model is able to reproduce reasonably well the empirical data, even if it tends to under-estimate production from dwarf fields in the 2000--2009 period (Fig.\  \ref{fig:oilModel}a).  The model allows also of forecasting future oil production trends, an exercise that leads to the prediction of the commercial exhaustion of Norwegian oil reserves before 2040 (Fig.\  \ref{fig:oilModel}b, black line).  It is also possible to build an alternative scenario where only the first resource, namely the giant oil field, is exploited.  This leads, understandably, to a lower production peak.  However, it is especially interesting to note both that the peak itself would have occurred about 10 years later than the real case and that the exhaustion of reserves would have been delayed by about 20 years (Fig.\  \ref{fig:oilModel}b, red line).  Framed in terms of sustainability, it is clear that using only the first resource would have led to a longer, and hence more sustainable, time span of resource use than the actual case of two resource exploitation.

\section{Model application 2: Antarctic whaling}\label{sec:whales}

While the first illustration of our model regarded a non-renewable resource---where it is not possible to achieve sustainability in a strict sense, but only to slow down the exploitation pace sufficiently to find adequate substitutes \citep{Daly1990}---in this section we apply it to a renewable resource, i.e.  a case where a sustainable equilibrium between human exploitation and resource regeneration rate is, at least theoretically, possible.

\begin{figure}[!t]%
\centering
\includegraphics[width=0.333\textwidth]{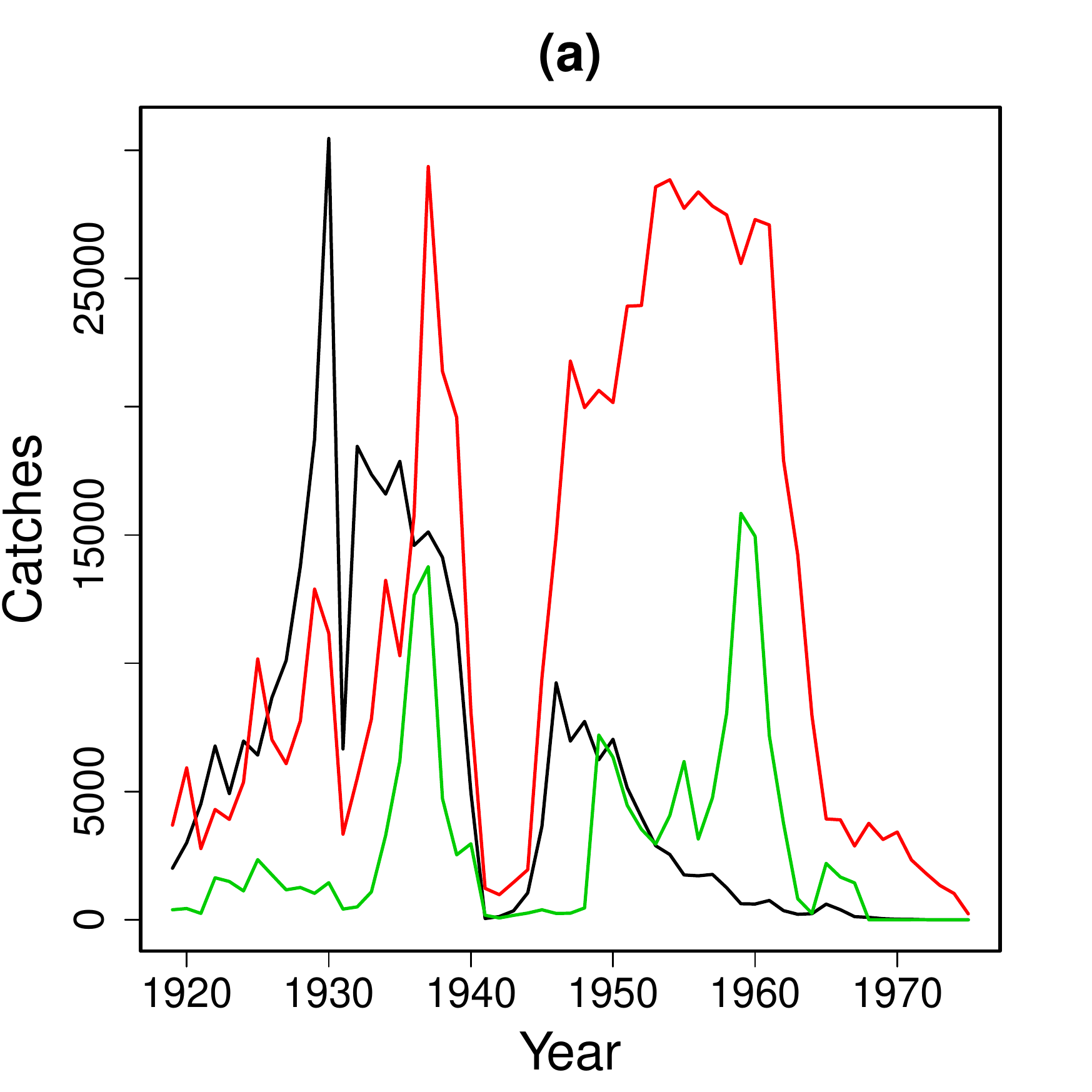}\hspace{1cm}\includegraphics[width=0.333\textwidth]{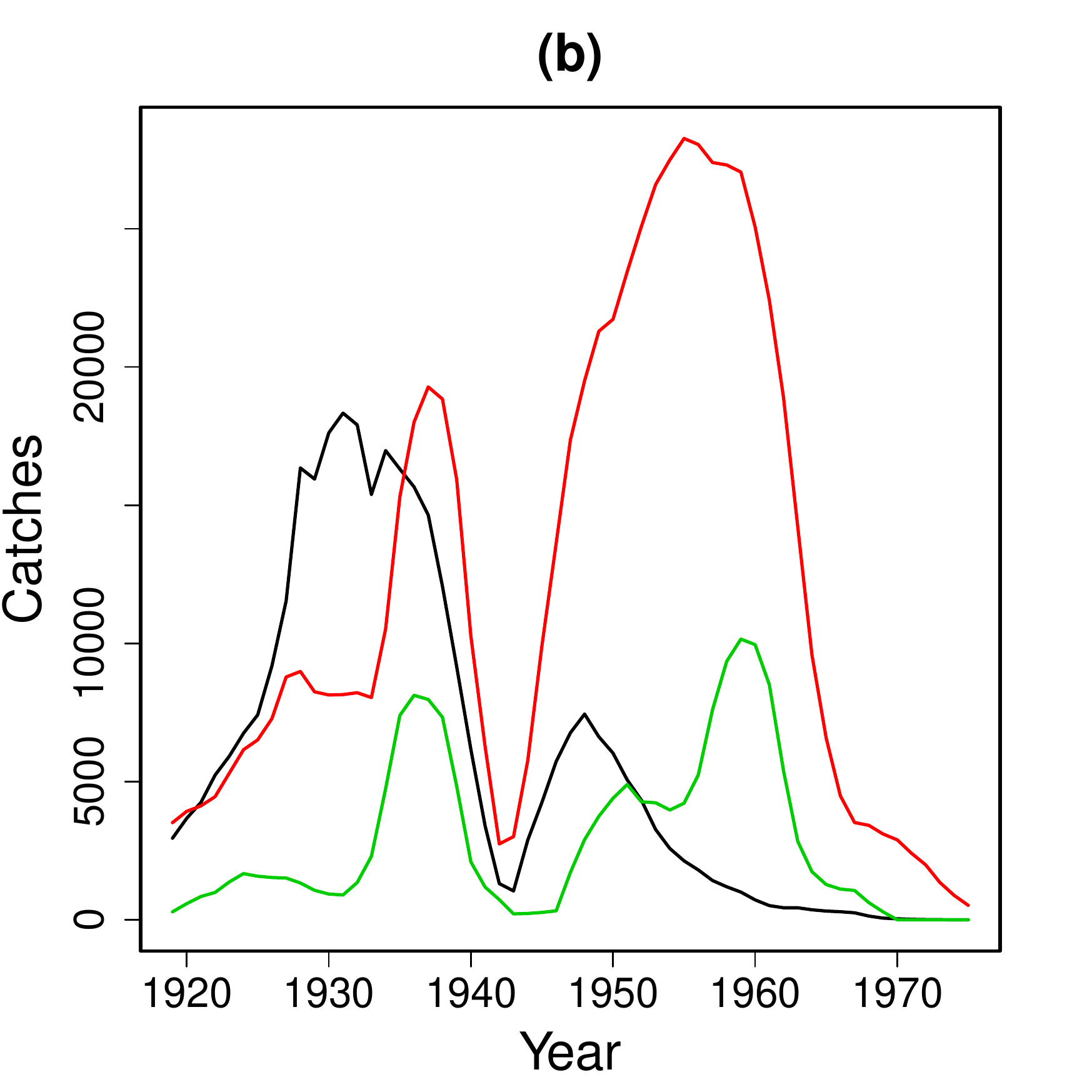}%
\caption{Blue, fin and humpback whale catches in the Antarctica 1919--1975.  The (a) panel presents the actual data, the (b) one shows five year mobile means on the same data.}%
\label{fig:whaleCatches}%
\end{figure}

Before the 20\textsuperscript{th} century, whaling interested mainly the \textit{Eubalaena glacialis} and the \textit{Balaena mysticetus}: relatively easy prey due to their slow swimming speed and to the fact that their dead bodies tend to float.  Due to excessive hunting, at the beginning of the 20\textsuperscript{th} century these species where almost commercially extinct in the North Atlantic.  However, technology improvements allowed for the first time the hunting of species belonging to the Balaenopteridae family \citep{Cherfas1988}.  After the First World War, Antarctic blue  whales (\emph{Balaenoptera  musculus}) became the main target of the new industry, with catches that reached a peak in the early 1930s and experienced a rapid decline subsequently.  With blue whales becoming rare, hunters turned to new targets.  From the 1930s to the 1960s---with the exclusion of the World War II period when whaling almost stopped---fin (\emph{Balaenoptera  physalus}) and humpback whales (\emph{Megaptera  novaeangliae}) replaced the blue whales as main hunt targets (Fig.\  \ref{fig:whaleCatches}a).  Finally, from the 1960s to the late 1970s, when  an international treaty banned  commercial hunting, most of catches were made up of sei (\emph{Balaenoptera  borealis}) and later minke whales (\emph{Balaenoptera  acutorostrata}) \citep{Branch2004,IWC2002,Mori2004}.  

A closer look at catch data shows a clear pattern: the most productive whales were exploited first, with a subsequent shift towards smaller species, with sperm whales (\emph{Physeter macrocephalus}) representing a partial exception due to the special use of their oil, e.g., in the US and USSR space rocket programs.  In terms of oil production, a single blue whale was indeed worth 2 fin, 2.5 humpbacks, 6 sei and approximatively 13 minke whales: a sequence that mimics almost exactly the shift of catch targets.  In other words, hunters first concentrated on the most valuable species, shifting progressively to less productive ones when the former target disappeared \citep{Clark1982,Cherfas1988,Schneider2004}.  

These considerations suggest that our model is well suited for application to the Antarctic whaling case.  Here $U$ represents the whaling industry (more precisely the whaling fleets operating in the Antarctic) and $R_1$, $R_2$, and $R_3$  the populations of, respectively, blue,  fin, and humpback whales.  We preferred to exclude from the model sei and minke whales, which were hunted later and under different market conditions (i.e., a stronger interest for whale meat and a reduced value of oil).  As time unit, we used the catching season and we evaluated all populations in terms of their economic value, choosing as unit the commercial value of a blue whale.  

We used the logistic growth function for whale populations, which implies a growth in condition of intra-specific competition.  The exclusion of the inter-specific competition followed from the considerations outlined in \citet{Mori2004}:  large baleen whales, such as humpback and fin, are distributed farther North during the Antarctic summer and are not as heavily dependent on krill as are blue whales.  This means that blue, fin and humpback whales have ecological niches which are not as closely overlapping as, for instances, blue and minke whales.  Inter-specific competition can hence be considered weak and need not to be modeled, at least in a first approximation.  
 
We considered that $e_1=e_2=e_3\equiv e$ could represent a reasonable assumption also in this case, since  whaling facilities could be used to catch any whale and investments did not depend from the specific targets.  Part of the reinvested sums were used to compensate the annual costs ($\mu U$ in our model), while the remaining resulted in an increasing of the capital $U$.  Given these considerations, model  (\ref{mod-gen}) applied to Antarctic whaling becomes:
\begin{equation} \label{mod-whale}
\begin{lcase}
	\dot{U}= &-\mu U + e\left(\frac{UR_1}{c_1+U+\alpha _1R_1} + \frac{UR_2}{c_2+U+\alpha _1R_1+\alpha _2R_2} + \right.  \\
	& \left.  + \frac{UR_3}{c_3+U+\alpha _1R_1+\alpha _2R_2+\alpha_3R_3}\right) \\
	\dot{R_1}= & r_1R_1\left(1- \frac{R_1}{K_1}\right) - \frac{UR_1}{c_1+U+\alpha _1R_1}\\
	\dot{R_2}= & r_2R_2\left(1- \frac{R_2}{K_2}\right) - \frac{UR_2}{c_2+U+\alpha _1R_1+\alpha _2R_2}\\
	\dot{R_3}= & r_3R_3\left(1- \frac{R_3}{K_3}\right) - \frac{UR_3}{c_3+U+\alpha _1R_1+\alpha _2R_2+\alpha_3R_3}
\end{lcase}
\end{equation}

The crucial issue in building a credible model is to assign realistic values to our parameters.  Since  precise cost data are available only after  WWII, we will use this period as a reference.  \citet{Clark1982} argued that the building cost of a complete fleet, composed by a factory ship plus usually 5--10 catchers,  ranged from 10  to over 20 million US Dollars.  The daily operating cost was 5000 USD per catcher, including fixed costs and depreciation.  \citet[Appendix]{Clark1982} reported a total of 2000 catcher-days per factory per year, but from other sources---including \citet{Cherfas1988}, \citet{Ponting1991} and  \citet[Main text]{Clark1982} themselves---a total of about 700 catcher-days per factory per year looks more reasonable.  This led us to estimate $\mu = 0.23$.

To estimate the parameters $\alpha_1$, $\alpha_2$ and $\alpha_3$, let us recall that $\alpha_i^{-1}$ represents the maximum  catch   per fleet, i.e., the catch  in the presence of an unlimited availability of the i-\textit{th} resource, see (\ref{enough}).  We assumed that  this condition held in the second decade of the 20\textsuperscript{th} century, at the very beginning of massive Antarctic whale exploitation.  Catch data from that period suggested an amount of 2000 whales per factory per year.  It seems reasonable to consider this figure as independent from the whale species, which implies that the catchability of the different species is about the same \citep{Clark1982}.  This number has to be multiplied for the economic value of a single whale expressed in the chosen unit.  The value of a single blue whale ranged from 7,000 USD in the European market to 11,000 USD in the Japanese one \citep{Clark1982}.  Using the intermediate value of 9000 USD, we estimated $\alpha _1 = 0.83$ and, recalling that a blue whale was worth 2 fins and 2.5 humpbacks, $\alpha_2  = 1.67$ and $\alpha_3 = 2.08$.   

It is reasonable to assume that the carrying capacity of the blue, fin and humpback whales populations---namely $K_1$, $K_2$ and $K_3$---was equal to the existing populations of pre-whaling times.  \citet{Branch2004} presented a reliable pre-whaling population estimates of Antarctic blue whales of 239,000 individuals while, according to \citet{Braham1984}, about 400,000 fin  and 100,000 humpback whales lived in the same area.  As before, these figures have to be multiplied by the economic value of each whale species expressed in the chosen unit.  We hence obtained $K_1 = 239,000$, $K_2 = 200,000$ and $K_3 = 40,000$.

\citet{Branch2004,Branch2006} and \citet{Matsuoka2006} provided estimates  for  $r_1$, $r_2$ and $r_3$, i.e., the annual growth rate of blue, fin and humpback whales.  According to these sources we set as referring values: $r_1=7.3\%$,  $r_2=10\%$ and $r_3=9.6\%$.  
 
>From (\ref{enough}) we see that $e/\alpha_i >\mu$ must hold for $i = 1,2$ because, in the real case, at least both blue and fin whales were capable of sustaining the whaling industry by themselves before the excessive reduction of their populations.  This provides the lower bound $e>0.38$, while the upper bound is obviously 1.  Unfortunately, it is not possible to derive a precise value of $e$ from the existing literature, even if we know that dividends arising from the whale industry were usually high, especially in the early period of Antarctic whaling, with dividends that where often higher than wages \citep{Clark1982}.  Therefore, it is likely that the value of $e$ should be significantly lower than the previous case.  We hence selected $e=0.4$.  

To estimate the $c_i$ values, we followed a strategy similar to the one described for the Norwegian oil case.  Note that the Second World War resulted in an almost complete stop of hunting operations (Fig.\  \ref{fig:whaleCatches}), although it did not allowed for a significant recovering of whales populations \citep{Branch2004}.  We hence excluded the corresponding years both from the model runs and from the catch dataset.  The original data being highly variable from year to year, instead of fitting directly the model on it, we preferred to use the corresponding five year mobile means (Fig.\  \ref{fig:whaleCatches}b) as target.  This allowed of avoiding an excessive influence of strong outliers (e.g., the over 30,000 blue whales caught in 1930) on our estimates.  This led to set the following values: $c_1=145,000$, $c_2=95,000$ and $c_3=1,650,000$.  

\begin{figure}[!t]%
\includegraphics[width=0.333\textwidth]{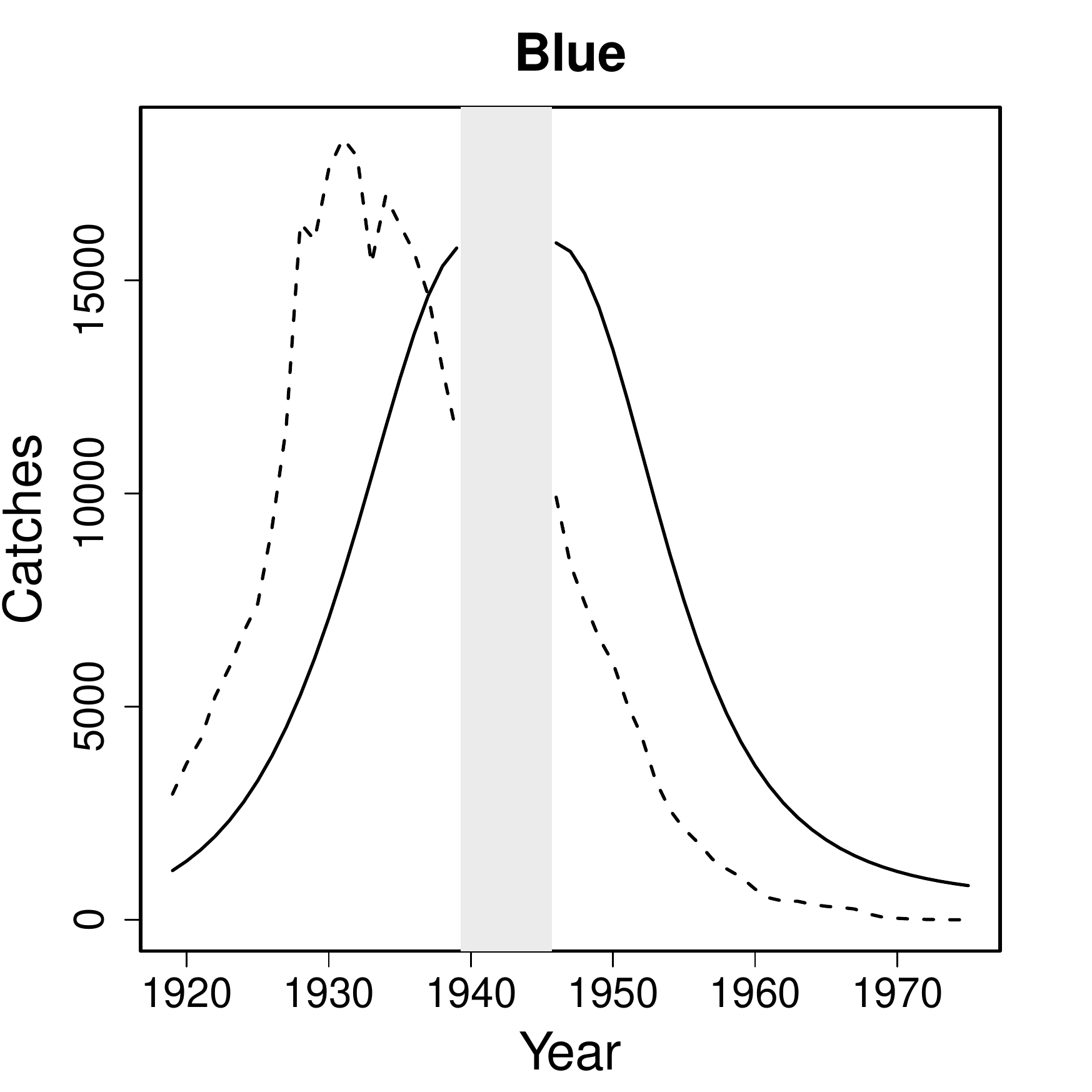}\hfill\includegraphics[width=0.333\textwidth]{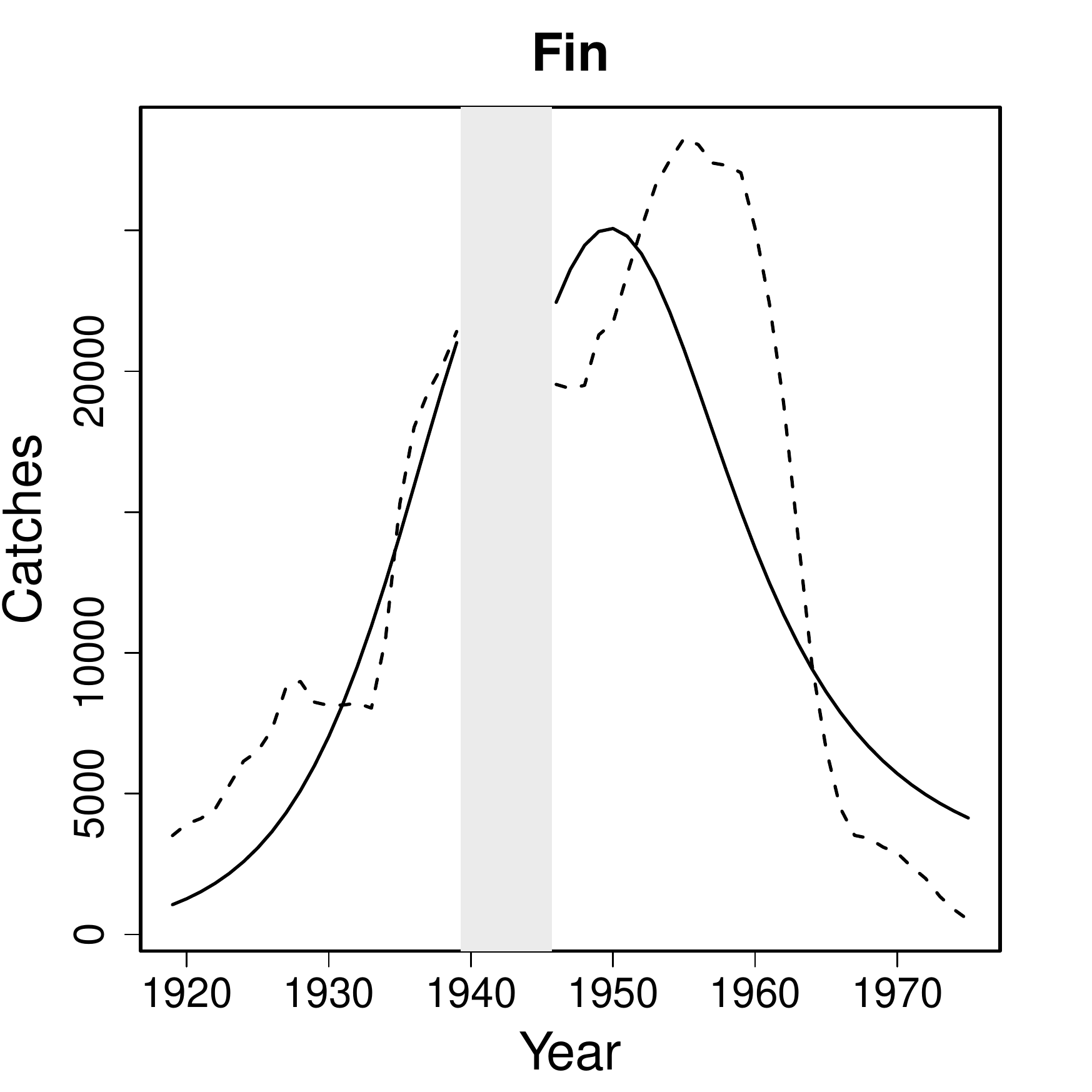}\hfill\includegraphics[width=0.333\textwidth]{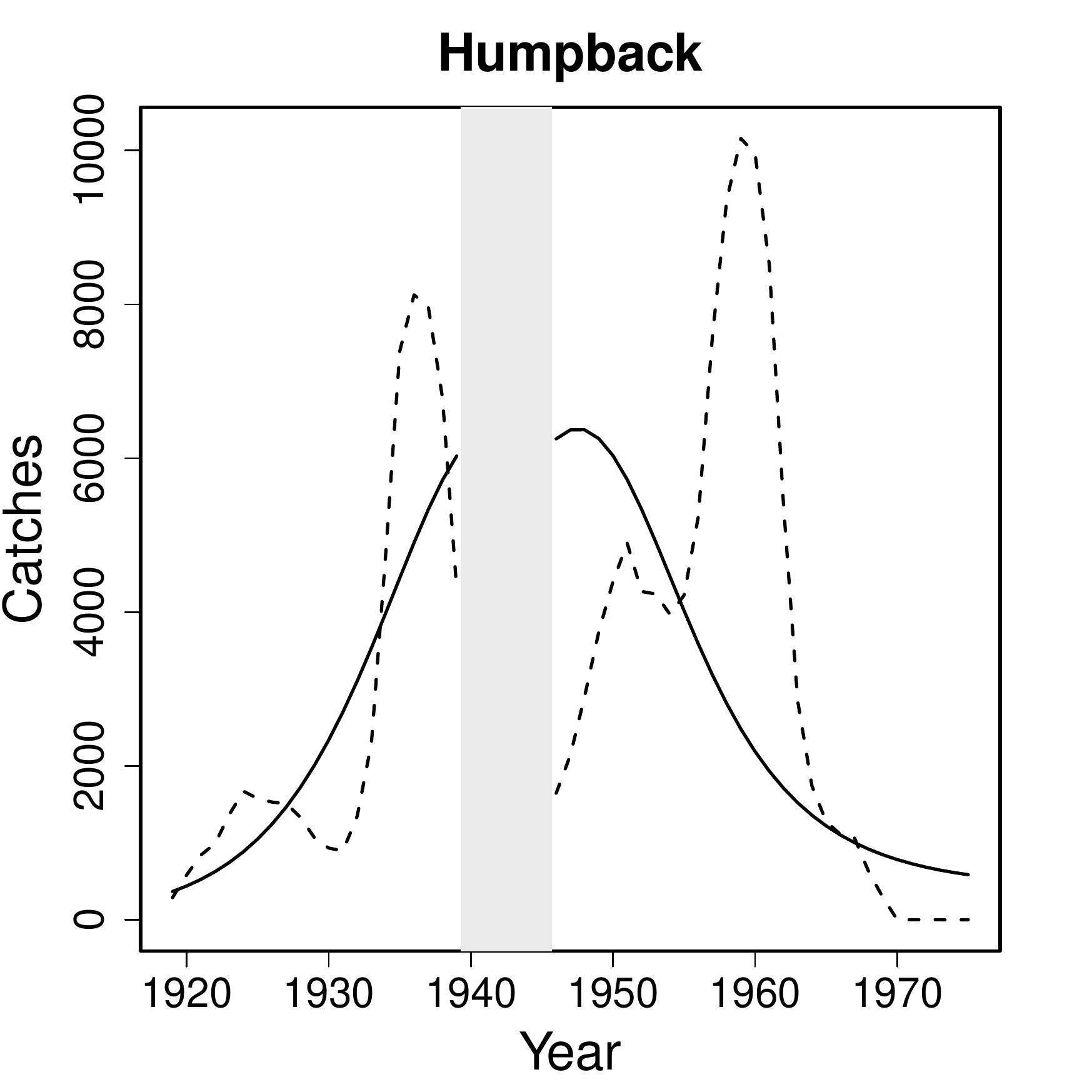}%
\caption{Simulated and real catch data (five year mobile means) for, respectively, blue, fin and humpback whales.  The vertical grey band corresponds to World War II period.}%
\label{fig:whaleModel}%
\end{figure}

Figure  \ref{fig:whaleModel} presents an overview of model output using the parameters described above.  Overall, our model is able to reproduce qualitatively the shift in the main hunting target that occurred in the decades under consideration, even if it tends to under-estimate the actual figures and the catch peak for blue whales is delayed.   This is probably due to the fact that other species not included in the model---notably sei and minke whales, along with sperm whales---represented important targets for the real whaling industry and fostered its growth at a level above the one that results from our model output \citep{Cherfas1988,Schneider2004}.  

\begin{figure}[!t]%
\centering
\includegraphics[width=0.333\textwidth]{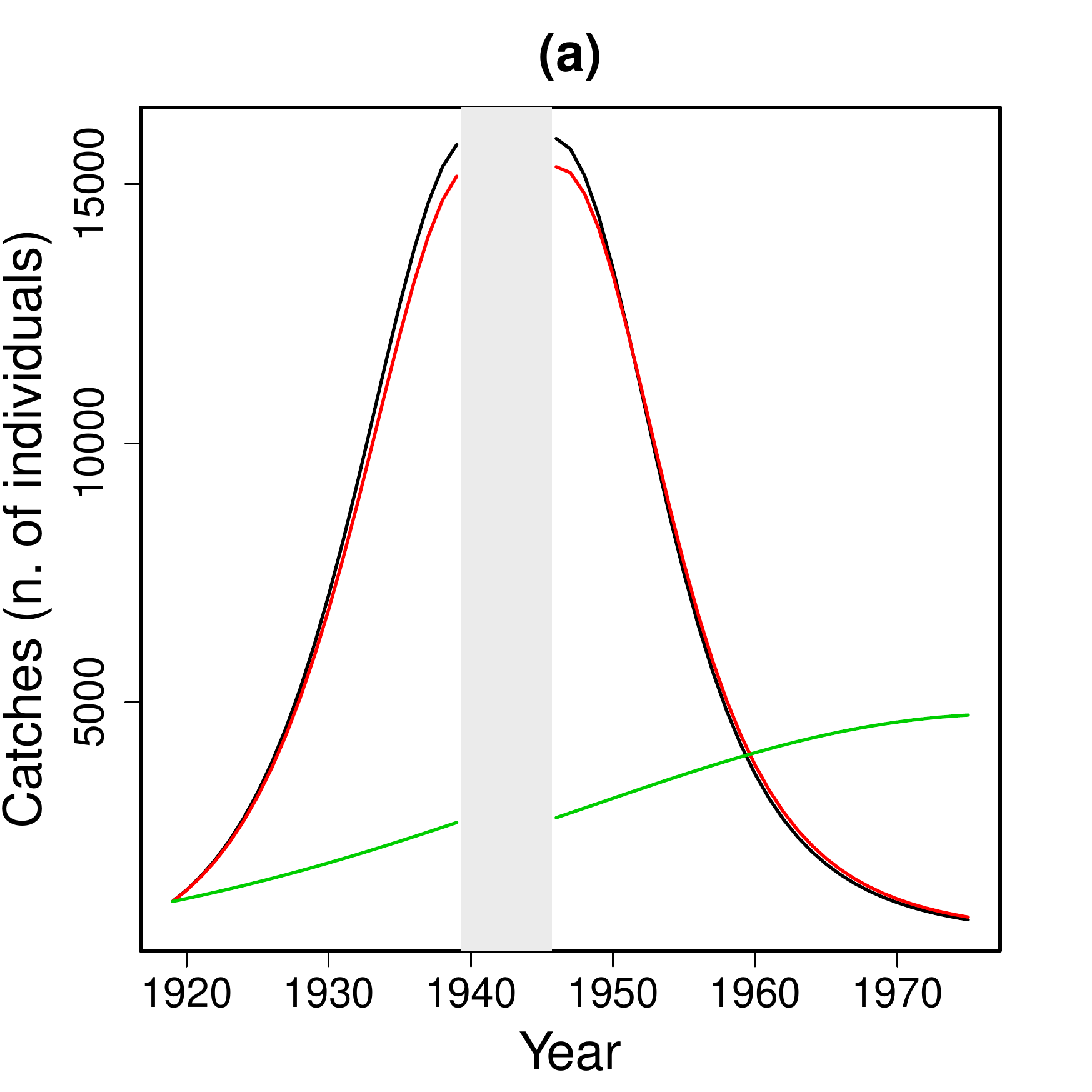}\hspace{1cm}\includegraphics[width=0.333\textwidth]{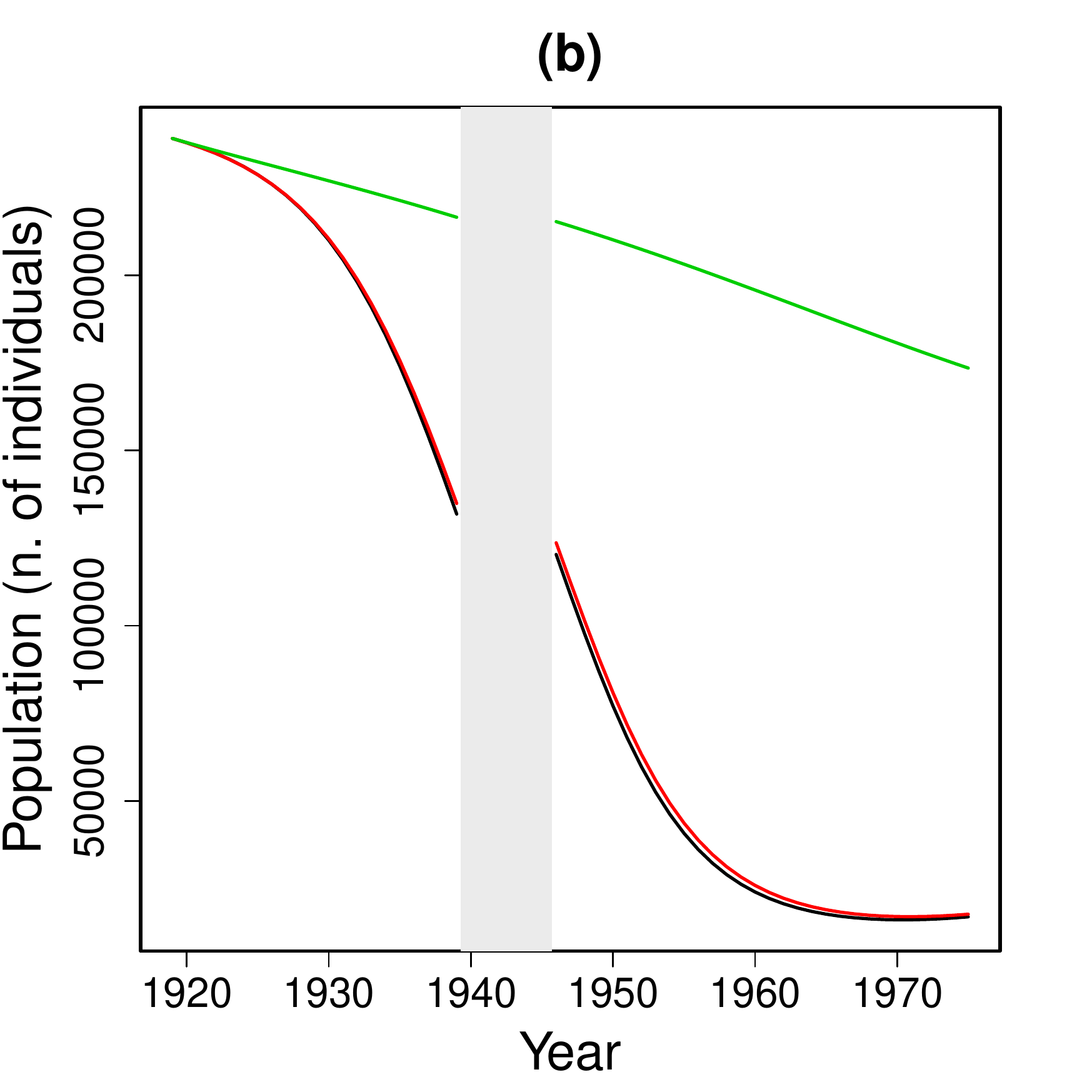}%
\caption{Blue whale catches (a) and population (b) in different runs on the model including: all three resource (black line), two resources (red line) and only one resource (green line).  The vertical grey band corresponds to the World War II period.}
\label{fig:whales123}%
\end{figure}

Even knowing that our model tends to under-estimate the impact on whaling on whales populations, we still can try to understand what could have happened if no switching to less productive resources had been available.  Figure  \ref{fig:whales123} plots the model output when including: (i) blue, fin and humpback whale (black line); (ii) blue and fin whales (red line); (iii) only blue whales.  The exclusion of the third resource proved not to be crucial in changing the model behavior, with only a modest shift of peak catches to the right and both trends following closely the dynamic of the real population of Antarctic blue whales  \citep{Branch2004}.  On the contrary, a stop of both fin and humpback hunt would have had rather dramatic consequences.  More specifically, the growth both of the catches and of the whaling industry would have been much slower and a large share of the original Antarctic blue whales population would have been maintained until the end of the period taken into consideration.  As before, the exploitation of a single species would hence have led to much more sustainable outcomes.

\section{Conclusions}\label{sec:conclusions}

Our results show that relying on multiple resources does not necessarily produce better outcomes than the exploitation of a single one.  Often, relying on more options leads indeed to a faster depletion of the preferred resource.  Moreover, abundance causes higher production peaks, followed by dramatic drops: a dynamic that ultimately reduces the possibilities of finding a sustainable equilibrium with the resource.

More specifically, our model shows that the final equilibrium, in terms of dimension of users' population, is similar or worse in the multiple resource case than in the single resource one.  Moreover, resources are exploited unsustainably  with all but the least preferred resource becoming severely depleted or destroyed.  In other words, under a large range of conditions, relying on more resources does not lead to a significantly higher welfare for users and simultaneously increases the environmental impact of their actions.

The two empirical illustrations show that this conclusion holds for both renewable and non-renewable resources.  This is especially clear for the Norwegian oil case.  The extra growth allowed by the exploitation of dwarf fields in addition to the giant ones, accelerated the  overall development rate of the oil industry.  This led, in turn, to a faster rate of consumption of oil reserves, with the effect that the production peak occurred about ten years before the single resource case and the final commercial exhaustion of the field has possibly been anticipated by twenty years.  Note also that the lower production peak in the case of single resource use would have required smaller changes for post-peak energy production, with a cheaper and easier to manage transition phase.  

The interpretation of the whaling case is less straightforward due to the larger number of elements not included in the model.  For simplicity, we only considered three whaling species---namely blue, fin and humpback whales---while pelagic whaling actually targeted also sei, minke and sperm whales.  This led us to under-estimate the growth of the whaling industry.  Nevertheless, our general conclusions did not change: had hunters relied only on blue whales, this would have led to a more sustainable equilibrium, with no commercial extinction of any species.  Note that this condition would be largely preferable to the actual one.  Currently, official hunting (although some pirate hunting still exist) is banned, even if at least two countries, namely Japan and Norway, use the  excuse of doing ``scientific research'' to maintain some level of commercial whaling.  However, this is possible only thanks to the abstention of all other whaling countries.  If commercial whaling were fully restored, besides its ecological impacts, it would indeed quickly become unprofitable for everybody \citep[see][]{Schneider2004}.

To sum up, consistently with the idea of the ``paradox of enrichment'', our work challenges the commonsense notion that relying on a larger array of resources is necessarily better.  This holds when resources are perfects substitutes and accurate management is at work, allowing for a switching between equally valuable alternatives  \citep[e.g.][]{Katsukawa2003}.  
However, in the real world where perfect substitutes rarely exists, the availability of several
resources often simply mean an unsustainable growth of the exploiting industry.  In this case, the achievement of a higher overshoot 
relative to the resource carrying capacity produces a subsequent deeper drop of the industry.  Careful management should hence take into account this risk by regulating the exploitation of \emph{each} of the resources under consideration, even when alternatives exists.  If this does not occur, the risk is simply to attain a worse equilibrium with the least preferred resource than the one that could have been obtained with the most preferred one.

\paragraph{Aknowledgements:}
The authors gratefully acknowledges comments and suggestions by Ignacio Monz\'{o}n and Ezio Venturino, along with the observations made by the participants during a seminar held at the Collegio Carlo Alberto in November 2010.

\bibliographystyle{authordate1}
\bibliography{c:/Documenti/Bibliografie/GB}

\end{document}